\documentclass[12pt]{iopart}

\usepackage{graphicx}  
\usepackage{bm} 

\expandafter\let\csname equation*\endcsname\relax
\expandafter\let\csname endequation*\endcsname\relax
\usepackage{amsmath}

\usepackage{verbatim}   
\usepackage{braket} 

\newcommand{\des}{\hat{a} }
\newcommand{\cre}{\ensuremath{\hat{a}^{\dagger}}}

\newcommand{\EJ}{\ensuremath{ E_{\text{J}}}}
\newcommand{\Lagr}{\mathcal{L}}

\begin{document}

\title{Multi-Photon Resonances in Josephson Junction-Cavity Circuits}

\author{Ben Lang and Andrew D. Armour}
\address{Centre for the Mathematics and Theoretical Physics of Quantum Non-Equilibrium Systems and School of Physics and Astronomy, University of Nottingham, Nottingham NG7 2RD, United Kingdom}
\ead{ben.lang@nottingham.ac.uk}

\begin{abstract}
We explore the dissipative dynamics of nonlinearly driven oscillator systems tuned to resonances where multiple excitations are generated. Such systems are readily realised in circuit QED systems combining Josephson junctions with a microwave cavity and a drive achieved either through flux or voltage bias. For resonances involving 3 or more photons the system undergoes a sequence of two closely spaced dynamical transitions (the first one discontinuous and the second continuous) as the driving is increased leading to steady states that form complex periodic structures in phase space. In the vicinity of the transitions the system displays interesting bistable behaviour: we find that coherent effects can lead to surprising oscillations in the weight of the different dynamical states in the steady state of the system with increasing drive. We show that the dynamics is well-described by a simple effective rate model with transitions between states localised at different points in the phase space crystal. The oscillations in the  weights of the dynamical states is reflected in corresponding oscillations in a time-scale that describes transitions between the states.  
\end{abstract}

\submitto{\NJP}
\maketitle

\section{Introduction}

Multi-photon resonances at which several photons can be created simultaneously via the down conversion of a high frequency pump are intrinsically highly nonlinear. Such processes are potentially very useful in a variety of contexts such as quantum error correction\,\cite{Mundhada_2019, Gottesman_2001}. However, multi-photon processes are also of intrinsic interest. They give rise to quantum states with a variety of novel properties  including higher-order squeezing\,\cite{Braunstein_1987,Chang_2020} and rich periodic structures in phase space, known as {\it{phase-space crystals}}\,\cite{Guo_2013, Guo_2020}.

In this paper we explore the dissipative dynamics of an oscillator that is driven nonlinearly so that a handful of photons (up to six) are generated at a time via a resonant process. Our work builds on several recent studies of 3-photon resonances\,\cite{Zhang_2017,Zhang_2019,Lorch_2019,Gosner_2020, Tadokoro_2020} as well as investigations of the properties of the eigenstates of the Hamiltonian that display periodic structures in the phase space, known as phase-space crystals, generated at very high-order resonances (typically more than ten photons)\,\cite{Guo_2013,Guo_2016}, though the latter focused mainly on the properties of closed systems.
We explore how the properties of the oscillator evolve as a function of the applied drive strength, the number of photons that are created together and the strength of the zero-point fluctuations. 

For resonances involving three or more photons the system undergoes a pair of closely spaced dissipative phase transitions: a first order one associated with the breaking of rotational symmetry in phase space and then a continuous one linked with a chiral symmetry of the system Hamiltonian. We find that novel features emerge for resonances involving more than three photons, with the system exhibiting a form of bistability in the vicinity of the dynamical transitions. In this regime the steady state consists of a mixture of states in the phase space with different amplitudes and phases. The weights of the different states oscillate as a function of the drive, leading to oscillations in the average occupation number of the oscillator. We examine the eigenspectrum of the Liouvillian of the system and construct a simple effective model for the slow dynamics of the system which reveals corresponding oscillations in the rate at which the system moves between coexisting high and low amplitude states.  We also analyse the eigenoperators of the Liouvillian which have a Bloch-like character in phase space, demonstrating how the concept of a phase space crystal can be applied in open systems.

Although multi-photon resonances can be found in a variety of different systems, the exceptionally strong non-linearities achieved in superconducting circuit devices make them especially suited to exploring this kind of physics\,\cite{Mundhada_2019,Chang_2020}. The specific nonlinearly driven oscillator model we analyse here can be realised by voltage-biasing a Josephson junction (JJ) in series with a microwave cavity. Photons are generated via inelastic tunnelling of Cooper pairs across the JJ and the system can be tuned to resonances where the creation of different numbers of photons is favoured by simply adjusting the bias voltage\,\cite{Hofheinz_2011,Armour_2013,Gramich_2013,Chen_2014,Leppakangas_2015,Cassidy_2017,Rolland_2019}. Furthermore, the resonances involving between 3 and 6 photons that we investigate are expected to be readily accessible with current devices. However, very similar effects can be achieved using slightly different architectures. For example, closely related Hamiltonians have been engineered by using a time-dependent flux bias instead\,\cite{Svensson_2017,Waltraut_2019}.  

The rest of this article is organised as follows. Section \ref{system} introduces our nonlinearly driven oscillator model in more detail. Then in section \ref{sec:fixedpoints} we analyse the classical fixed points of the system which provide a framework for analysing the quantum dynamics. We then examine the properties of the full (quantum) steady-state in section \ref{sec:steadystate}, looking in particular at how it depends on the drive strength and the number of photons generated. Next, in section \ref{sec:eigspec} we explore the eigenspectrum and corresponding eigenoperators of the Liouvillian of the system. This allows us to go on to develop a simple description of the dynamics in section \ref{sec:effective}, as a small set of dynamical states linked by transition rates between them. We summarise and conclude in section \ref{sec:conclude} and further details for certain aspects of the work are provided in Appendices. 

\section{Non-Linearly Driven Oscillator Model}
\label{system}

\begin{figure}[h!]
\centering
\includegraphics[scale=0.5]{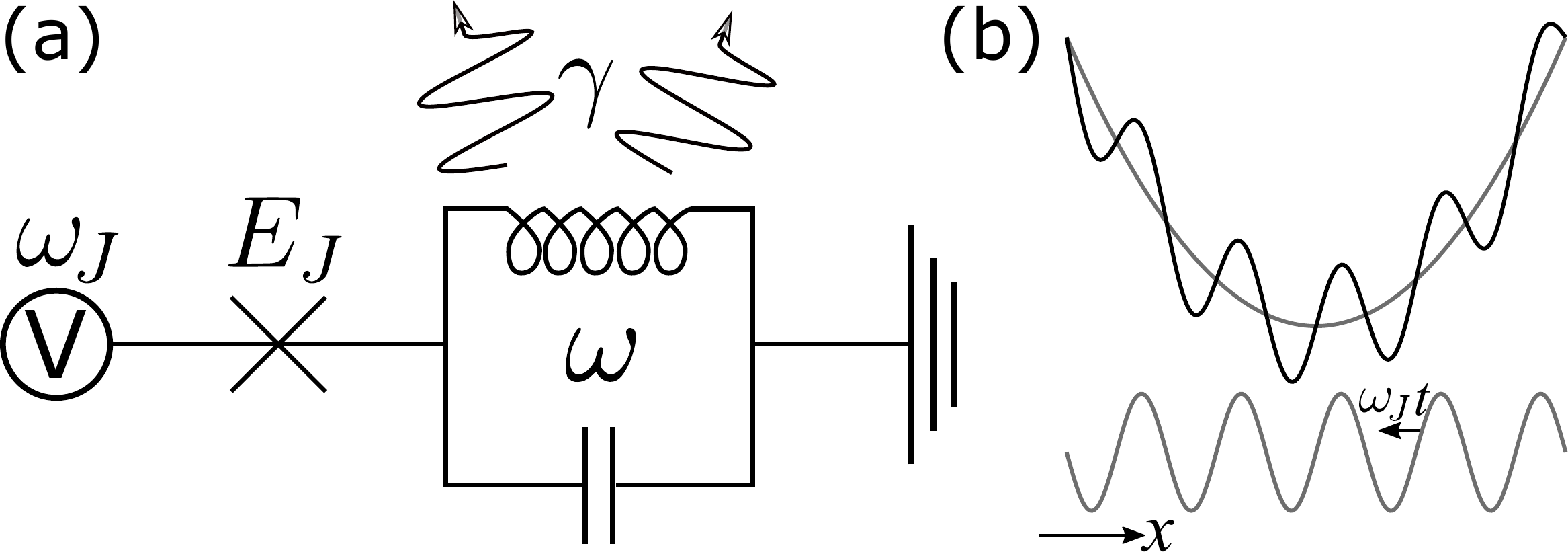}
\caption{(a) Schematic diagram of the circuit realisation of the model system we consider. An  $LC$ oscillator with frequency $\omega=\sqrt{1/LC}$  is driven at a frequency $\omega_J=2eV/\hbar$ by a voltage-biased Josephson junction, characterised by the Josephson energy $\EJ$. The oscillator is assumed to be damped at a rate $\gamma$. (b) Sketch of the time-dependent potential in the Hamiltonian. It consists of a harmonic quadratic part and a moving sinusoidal potential\,\cite{Meister_2015}.}
\label{circuit_diagram}
\end{figure}

The model circuit we consider is shown in figure \ref{circuit_diagram}, it consists of a series combination of a JJ and cavity described by a single LC oscillator with frequency $\omega_0=1/\sqrt{LC}$  (other cavity modes are assumed to be far detuned from resonance), though it could equally be realised with a JJ coupled to a lumped element LC oscillator\,\cite{Rolland_2019}. The Hamiltonian of the system is given by \cite{Armour_2013,Gramich_2013}
\begin{equation}
\hat{H}(t) = \hbar \omega_{\text{0}} \cre \des -  \EJ  \cos\left[ \omega_{\text{J}} t + \Delta (\cre + \des) \right],
\label{time_Hamiltonian}
\end{equation}
where  $\des$ is the lowering operator of the mode, $\EJ$ is the Josephson energy of the junction,   $\omega_{\text{J}} = 2 e V / \hbar$ is the Josephson frequency set by the bias voltage and $\Delta = ( 2 e^2 \sqrt{L/C}/\hbar )^{1/2}$. This Hamiltonian can also be regarded as describing a particle in a quadratic potential perturbed by a travelling wave\,\cite{Meister_2015}, and as a result the same Hamiltonian emerges in the study of cold atoms in time-varying optical traps\,\cite{Guo_2016}.

We consider the case where the voltage is tuned close to resonances where the inelastic tunnelling of a Cooper-pair can generate $p$-photons in the mode, so that $\omega_J\simeq p\omega_{\text{0}}$. Moving to a frame that rotates at a frequency $\omega_J/p$, and making a rotating wave approximation (RWA)\,\cite{Armour_2013} leads to the Hamiltonian
\begin{equation}
\hat{H}_{\text{RWA}} = \hbar \delta \cre \des - \frac{ \rmi^p \EJ \rme^{-\Delta^2/2}}{2} : [ (\des)^p + (-\cre)^p] \frac{J_p(2\Delta \sqrt{\cre \des })}{(\cre \des)^{p/2}}:, \label{rwaham}
\end{equation}
where $\delta = \omega_{\text{0}} - \omega_J/p$ is the detuning from resonance,  the colons, $: :$, indicate normal ordering and $J_p(x)$ is a Bessel function: $J_p(x) = \Sigma_n (-1)^{n} (x/2)^{p+2n} / n! (n+p)!$. Although the RWA approximation has been found to be a very good approximation for a wide range of parameters in this system\,\cite{Kubala_2015,Rolland_2019}, we have explicitly checked that it also works well for the parameter regimes we explore here, see \ref{Appendix: RWA}.

The RWA Hamiltonian has two important symmetries\,\cite{Guo_2016}. First, it commutes with the operator $\hat{r} = \exp( \rmi \cre \des (2\pi/p) )$, which follows from the batch creation of $p$-photons. In phase space this commutation manifests as discrete rotational symmetry\,\cite{Guo_2016,Zhang_2019}. This is the main ingredient of phase space crystals:  eigenstates of the Hamiltonian inherit this symmetry, which permits a description in terms of Bloch modes\,\cite{Guo_2013,Guo_2016,Zhang_2017}. The second symmetry is that the operator $\hat{c} = \hat{r}^{1/2}$ anti-commutes with the Hamiltonain, $\hat{c} \hat{H}_{\text{RWA}} = -\hat{H}_{\text{RWA}} \hat{c}$. This is \emph{chiral} symmetry, half-period rotation leading to a sign change \cite{Guo_2016}.

We include the effects of photon losses from the cavity together with the coherent evolution of the system via a Lindblad master equation, assuming  a zero-temperature environment for simplicity
\begin{equation}
\dot{\rho} = \Lagr(\rho) =  -\frac{i}{\hbar}[H_{\text{RWA}}, \rho] + \frac{\gamma}{2} ( 2 \des \rho \cre  - \cre \des\rho - \rho \cre \des ),
\label{Master_equation}
\end{equation}
with $\gamma$ the loss rate. 
In fact, the fixed bias voltage we have assumed is an idealisation: in practice the voltage across the JJ and the cavity is not completely fixed, but subject to fluctuations due, e.g.\, to the presence of additional impedances in the circuit\,\cite{Gramich_2013, Wang_2017}. Here we concentrate on the minimal description of the system dynamics (\ref{Master_equation}),  but our approach can be extended to account for the effects of voltage fluctuations and we consider how they affect our results in \ref{Appendix:volt noise}.

The discrete rotational symmetry of the RWA Hamiltonian (though not the chiral symmetry) is preserved in the master equation\,\cite{Minganti_2018}. In this case, $[\mathcal{L},\mathcal{R}]=0$ where $\mathcal{R}\,\cdot=\hat{r}\cdot\hat{r}^{-1}$. This symmetry does not give rise to a conserved charge as a conserved charge requires the unmet, stronger condition that the collapse operator, $\des$, commutes with $\hat{r}$ \cite{Albert_2014}.

Within the master-equation description, only two timescales remain when the voltage is tuned to resonance so that $\delta=0$. Their ratio, $\EJ/\hbar\gamma$, sets the strength of the driving compared to that of the losses. The value of $\EJ$ can be tuned in-situ by using a slightly more complex SQUID set-up\,\cite{Armour_2013} which acts as an effective single junction with a Josephson energy that can be tuned via an applied flux. The quantity $\Delta$ determines the strength of the quantum fluctuations, it measures the magnitude of the zero-point fluctuations in the oscillator flux in units of the flux quantum. Whilst $\Delta$ is fixed within a given experiment, recent developments in device engineering mean that it can be varied over a relatively wide range\,\cite{Rolland_2019}: from $\Delta\ll 1$ up to $\Delta\sim 1$. Looking at the Hamiltonian (\ref{rwaham}), we see that $\Delta$ mediates the strength of the Bessel function term. This describes how the presence of photons in the oscillator affect the creation of further photons: for $\Delta\sim 1$ this becomes significant even at the level of a single photon. The other parameter that appears is $p$, the number of photons generated at a time at the resonance with $\omega_J=p\omega_0$.

\section{Classical Fixed Points}
\label{sec:fixedpoints}
Before looking at the quantum dynamics of the system we examine the corresponding classical dynamics. Identifying the classical fixed points of the system, together with the sequence of bifurcations that arise as the driving strength is increased progressively, provides a very useful framework for analysing the open quantum dynamics. 

Equations of motion for a complex classical amplitude $\alpha$ can be obtained from the master equation  (\ref{Master_equation}) by simply making the ansatz that the system is in a coherent state\,\cite{Morley_2019} ($\rho\rightarrow |\alpha\rangle\langle\alpha|$). A more sophisticated semi-classical approximation can be developed by transforming the master equation into an equation of motion for the Wigner distribution (WD) of the density operator, $W(\alpha,\alpha^*)$, and dropping derivatives beyond second-order which leads to a Fokker-Planck equation (see \ref{Appendix:Wigner transform} for details). Although more involved, this route provides a more systematic approach to understanding the classical limit of  the quantum dynamics\,\cite{Lorch_2019}. 

The ways in which the fixed points evolve with drive strength and the bifurcations that emerge for different values of $p$ are necessarily constrained by the corresponding symmetries. For $p=1$ the fixed point which is at zero-amplitude ($A=|\alpha|=0$) for zero drive moves away from the origin smoothly with increasing drive (higher \EJ). However, for $p=2$ the $\pi$-rotational symmetry means the fixed point cannot move from the origin. Instead, at a particular value of \EJ, the origin loses stability\,\cite{Armour_2013}, with the simultaneous emergence of two stable fixed points in opposite directions---a pitchfork bifurcation\,\cite{Strogatz_book}.

For $p>2$ there is always a stable point located at zero amplitude which we call the Dark Point (DP). It cannot be displaced from the origin or undergo bifurcations while maintaining $3$-fold or higher rotational symmetry\footnote{This argument is quite general and applies to other damped systems driven at an overtone above 2 \cite{Svensson_2017, Zhang_2017, Tadokoro_2020}}. However the DP is not always the only stable point in these systems. 

For $p\ge 3$,  a set of $p$ new stable points are generated (together with $p$ corresponding saddle points), with a common amplitude $A_r$ and phases that differ by $2\pi/p$, via a set of saddle node bifurcations. The bifurcations occur at a threshold drive $\EJ^{(T)} = (\hbar\gamma A_r^2 \rme^{\Delta^2/2})/[p J_p(2\Delta A_r)]$ with the amplitude given by the smallest nonzero solution to $\Delta A_r J^{'}_{p}(2\Delta A_r) = J_{p}(2\Delta A_r)$ (see \ref{Appendix: Bifurcations} for details).
We call these stable points Bright Points (BP), and it is a key characteristic of the multiphoton resonances with $p\ge 3$ that these  points always coexist with the stable DP. The values of both $\EJ^{(T)}$ and $A_r$ increase with $p$, but whilst $A_r$ is proportional to $1/\Delta$, $\EJ^{(T)}$ instead scales as $\exp(\Delta^2/2)/\Delta^2$.

As $\EJ$ is increased beyond $\EJ^{(T)}$, the BPs migrate to higher amplitudes (with the corresponding saddle points moving to lower amplitudes) and eventually a second set of bifurcations take place. In this case pitchfork bifurcations occur as each BP splits into three, two stable points that remain locked at a fixed amplitude and thereafter evolve only in phase with increasing drive, and an unstable (saddle) point that continues to increase in amplitude. In contrast to the saddle node bifurcations, these bifurcations occur for all $p$ values\,\cite{Armour_2013,Armour_2017}. 

\begin{figure}[t!]
\centering
\includegraphics[scale=0.38]{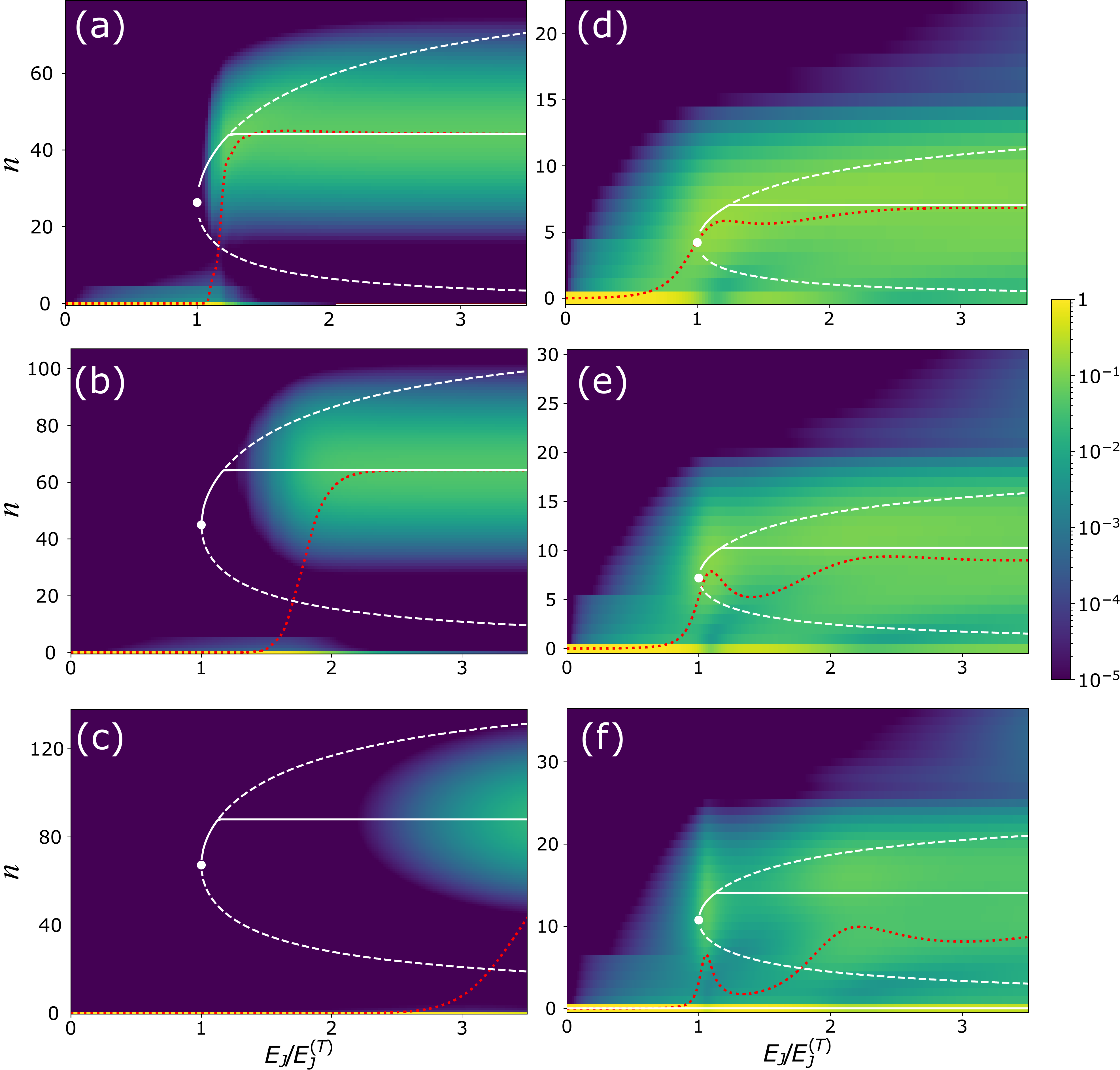}
\caption{Steady state behaviour as a function of $\EJ$. The probability of each photon number, $P_n=\langle n|\rho|n\rangle$ is plotted on a log-colourscale, the expectation value of photon number is indicated by the red-dashed line. Overlaid in white are the squared amplitudes of the fixed points, stable/unstable with solid/dashed lines. $\Delta =0.4$ for the first column (a,b,c) and $\Delta=1.0$ for the second (d,e,f), while $p$ takes the values 4 (a,d), 5 (b,e) and 6 (c,f).}
\label{photon_numbers}
\end{figure}

\section{Steady State Properties}
\label{sec:steadystate}
The strong nonlinearity of the multi-photon resonances mean that numerical methods\footnote{We used the QuTiP package\,\cite{QUTIP} for this task.} must be used to explore the quantum properties of the system.  The first question that we address is how the steady-state properties of the system, in particular the occupation number $\langle\hat{n}\rangle=\langle \hat{a}^{\dagger}\hat{a}\rangle$, behave and how this compares to the underlying structure of classical fixed points and bifurcations.  Then in the next section we investigate the key dynamical time-scales in the problem by calculating the spectral properties of the Liouvillian. 

Figure \ref{photon_numbers} shows the photon-number distributions of the steady states as a function of the drive strength, $\EJ$, for $p=4,5,6$ and  $\Delta=0.4$ and $\Delta=1.0$. The plots also show the behaviour of $\langle\hat{n}\rangle$ and the classical fixed points. 

The evolution of the classical fixed points with $\EJ$ follows a very similar pattern for different values of $p$ and $\Delta$. In each case, the first bifurcation occurs at $\EJ^{(T)}$, leading to the new stable BPs appearing well away from zero, followed by the second bifurcation after which the BP amplitude remains fixed. This fixed amplitude is given by\,\cite{Armour_2013} $A_p=z_p/(2 \Delta)$ with $z_p$ given by the first zero of $J_p'(z_p)$, so that $z_p=5.318,6.416,7.501$ for $p=4,5,6$. Since for all but the largest $\Delta$ and $p$ values, the average occupation numbers match up well with the prediction of the fixed point amplitudes in the limit of large $\EJ$, this provides a simple measure of how the occupation numbers in the problem increase with $p$, but decrease with increasing $\Delta$.  Another important effect of increasing $p$ is that it brings the two bifurcations closer together (indeed they merge in the infinite-$p$ limit). 

Although the behaviour of the average occupation number comes close to the corresponding classical fixed point for large enough $\EJ$ in almost all the plots shown in figure \ref{photon_numbers}, the differences are much more marked at lower drive strengths. In particular, the point at which the system displays a threshold, marked by a rapid rise in $\langle \hat{n}\rangle$ differs progressively from the classical prediction $\EJ^{(T)}$ as $p$ is increased for $\Delta=0.4$. For $\Delta=1$ the behaviour is qualitatively different with $\langle \hat{n}\rangle$ developing a series of peaks as $\EJ$ is increased which become more pronounced at higher $p$\footnote{We did not find any clear signatures of these oscillations for $p<4$.}. These oscillations are remarkable in that they imply that increasing the drive (or equivalently reducing the damping rate) can lead in places to a reduction of the occupation number.

The photon number distributions shown in figure \ref{photon_numbers} reveal significant bimodality over a broad range of drive strengths. In each case, a peak is always present for zero photons with a second peak emerging around a value associated with the corresponding classical fixed point (i.e.\ the BP). For $\Delta=0.4$ the peaks are well separated, becoming more distinct with increasing $p$ and the weight of the distribution shifts progressively away from the zero-photon peak as the drive is increased. In contrast, for $\Delta=1.0$ the peaks are much less well separated and we see that the oscillations in $\langle \hat{n}\rangle$ involve oscillations in the weight of the distribution back and forth between the two peaks.

\begin{figure}[h!]
\centering
\includegraphics[scale=0.6]{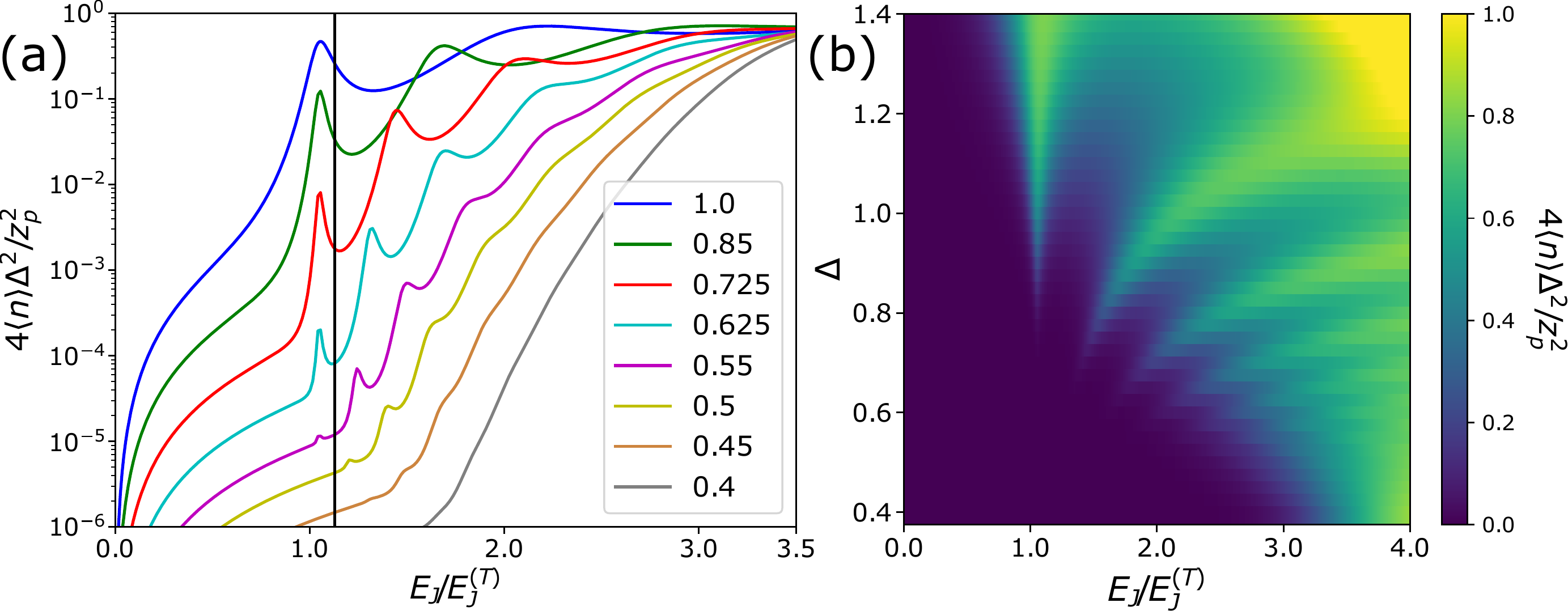}
\caption{(a) $\langle \hat{n} \rangle$ as a function of $\EJ$ for various $\Delta$ and $p=6$, log-scale. The photon numbers are scaled using the value at which the BP amplitude saturates, $z_p / (2 \Delta)$, so that the range is between zero and unity for all $\Delta$. (b) Behaviour of $\langle \hat{n} \rangle$ as a function of both $\Delta$ and $\EJ$ (with a linear colour-scale).}
\label{Log_N}
\end{figure}

We trace the evolution of the oscillations in $\langle\hat{n}\rangle$ as a function of $\EJ$ and $\Delta$ for $p=6$ in figure \ref{Log_N}. This shows that the oscillations form a series of curving resonances in the $\EJ$-$\Delta$ plane, developing for $\Delta \ge 0.45$, then becoming less distinct for $\Delta > 1$, suggesting that the peaks in the underlying distribution need to be neither too far apart,  nor too close together, for the oscillations to be clearly pronounced. For all values of $\Delta$ the first of the peaks occurs at the same $\EJ / \EJ^{(T)}$, between the two classical bifurcations, and the peaks eventually disappear for large enough drive strengths.

\begin{figure}[h!]
\centering
\includegraphics[scale=0.5]{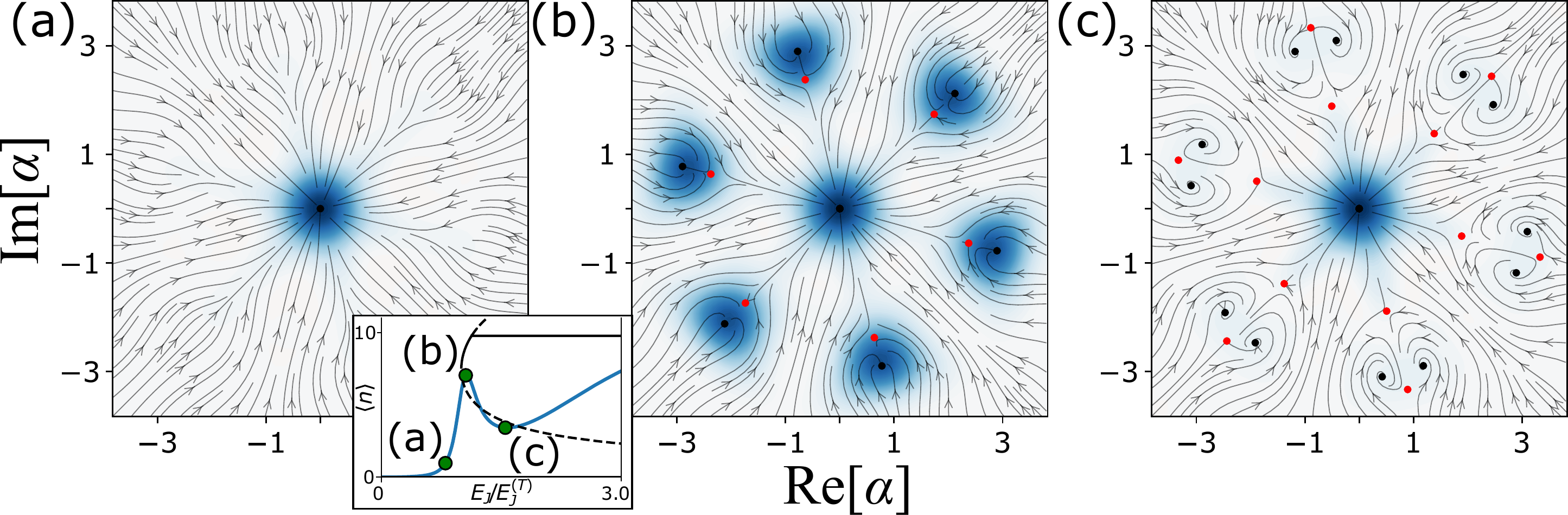}
\caption{Wigner densities of the steady state. Overlaid are streamlines of the coherent state drift terms (arrows), and the stable/unstable fixed points in black/red. The panels show $p=6$, $\Delta = 1.2$ with $\EJ / \EJ^{(T)} \approx [0.8, 1.1, 1.6]$ respectively. The WD is proportional to the intensity of the blue colour. Inset:  $\langle n \rangle$  as a function of $\EJ$, dots indicate the values used in the main plots. }
\label{Wigners}
\end{figure}

Further insights can be obtained by looking at the evolution of the system in phase space using the WD of the steady states, as shown in figure \ref{Wigners}. The WDs are shown together with the classical fixed points and the flow streamlines (classical trajectories) for a sequence of increasing $\EJ$ values\footnote{Note that the WDs are almost entirely positive. Panels (a), (b), (c) of figure \ref{Wigners} have negativity volumes\,\cite{Kenfack_2004}, of $1\times 10^{-2}$, $1.2 \times 10^{-4}$ and $8.7 \times 10^{-2}$ respectively, too faint to display in the figure itself.}. In each case the steady state consists of `blobs' localised about the classical attractors\,\cite{Armour_2017,Bartolo_2016,Roberts_2020}. The drop in photon population as $\EJ$ rises is clearly visible in the almost total drop in the magnitude of the WD around the BPs between panels (b) and (c).

The oscillations in $\langle \hat{n} \rangle$ are a feature of the full coherent quantum dynamics of the system. We could find no trace of them using a semi-classical description in terms of the Fokker-Planck equation for the Wigner function (obtained by dropping higher order derivative terms in the equation of motion for the WD) which incorporates diffusion on top of the classical dynamics (see \ref{Appendix:Wigner transform} for details). Furthermore, the coherences in the number state basis play a crucial role. Introducing number-dephasing terms into the master equation to mimic the effects of voltage noise that arises in experiments\,\cite{Souquet_2016,Gramich_2013,Rolland_2019} with voltage bias JJ-cavity systems, we find that the sharp resonances in $\langle \hat{n} \rangle$ are progressively washed out as the strength of the dephasing is increased (see \ref{Appendix:volt noise}).

\section{Liouvillian Eigenspectrum}
\label{sec:eigspec}

The change of the steady state from one localised at the DP to one localised about the BPs is surprisingly complex, associated as it is with an extended region of bimodality in which the weights of the distribution around the different points can oscillate as a function of the drive. To understand more about this behaviour we turn now to the properties of the Liouvillian super-operator, defined in (\ref{Master_equation}). 

The Liouvillian ($\Lagr$) transforms one operator into another\,\cite{Minganti_2018}. It has a set of (not-necessarily Hermitian) eigenoperators, $\rho_n$ which satisfy:
\begin{equation}
\Lagr \rho_n =  \lambda_n \rho_n,
\label{Liou}
\end{equation}
with $\lambda_n$ the corresponding eigenvalue. As the evolution of the density operator is determined by the Liouvillian, $\rme^{\Lagr t}$, these eigenoperators all have a simple time evolution: $\rho_n(t) = \rme^{\lambda_n t} \rho_n(0)$. The steady state is the eigenoperator with eigenvalue zero. There is only ever one such state for systems like ours with annihilation-operator dissipation  \cite{Schirmer_2010}. All other eigenvalues have negative real part, so that an arbitrary initial state expressed as a linear combination of eigenoperators converges towards the steady state under time-evolution\,\cite{Macieszczak_2016,Minganti_2018}. Time evolution conserves trace and quasiprobability, thus the steady state can be normalised so that its density operator has trace 1 and its WD has integral 1. In contrast the transients all require a trace/integral of 0.

Dissipative phase transitions\,\cite{Kessler_2012,Macieszczak_2016,Minganti_2018} are associated with the emergence of one or more very slow time-scales which arise when parameters of the system are tuned, leading to a sharp narrowing of the Liouvillian gap, given by $|{\rm{Re}}[\lambda_1]|$ where the eigenvalue $\lambda_1$ is the (non-zero) eigenvalue whose real part is least negative. Formally, a phase transition occurs in a thermodynamic limit, associated with a divergence in occupation number for nonlinear oscillator systems\,\cite{Minganti_2018}, where the Liouvillian gap vanishes.

The eigenoperators, $\rho_n$, inherit the symmetries of the Liouvillian. In our case, there is a discrete rotational symmetry in phase space, described by the rotation super-operator, $\mathcal{R}$. Since $[\mathcal{R},\Lagr]=0$,  eigenoperators of $\Lagr$ will also be eigenoperators of $\mathcal{R}$ with $\mathcal{R} \rho_n = \rme^{\rmi k_n} \rho_n$, with $k_n$ an integer multiple of $2 \pi/p$. We will see that this symmetry plays a crucial role in applying the concepts of phase space crystals in an open system\footnote{The rotational invariance of the collapse operators means that the Liouvillian does not inherit the chiral symmetry present in the Hamiltonian.}. Seen in the lab frame this symmetry corresponds to time-translation symmetry \cite{Minganti_2020}.

In the following we start by analysing the behaviour of the eigenvalues of the Liouvillian and the connections to dissipative phase transitions in detail. We also uncover clear connections between the behaviour of the eigenvalues and the oscillations in the occupation number discussed in the previous section. We then go on to explore the properties of the corresponding eigenoperators. 

\begin{figure}[h!]
\centering
\includegraphics[scale=0.5]{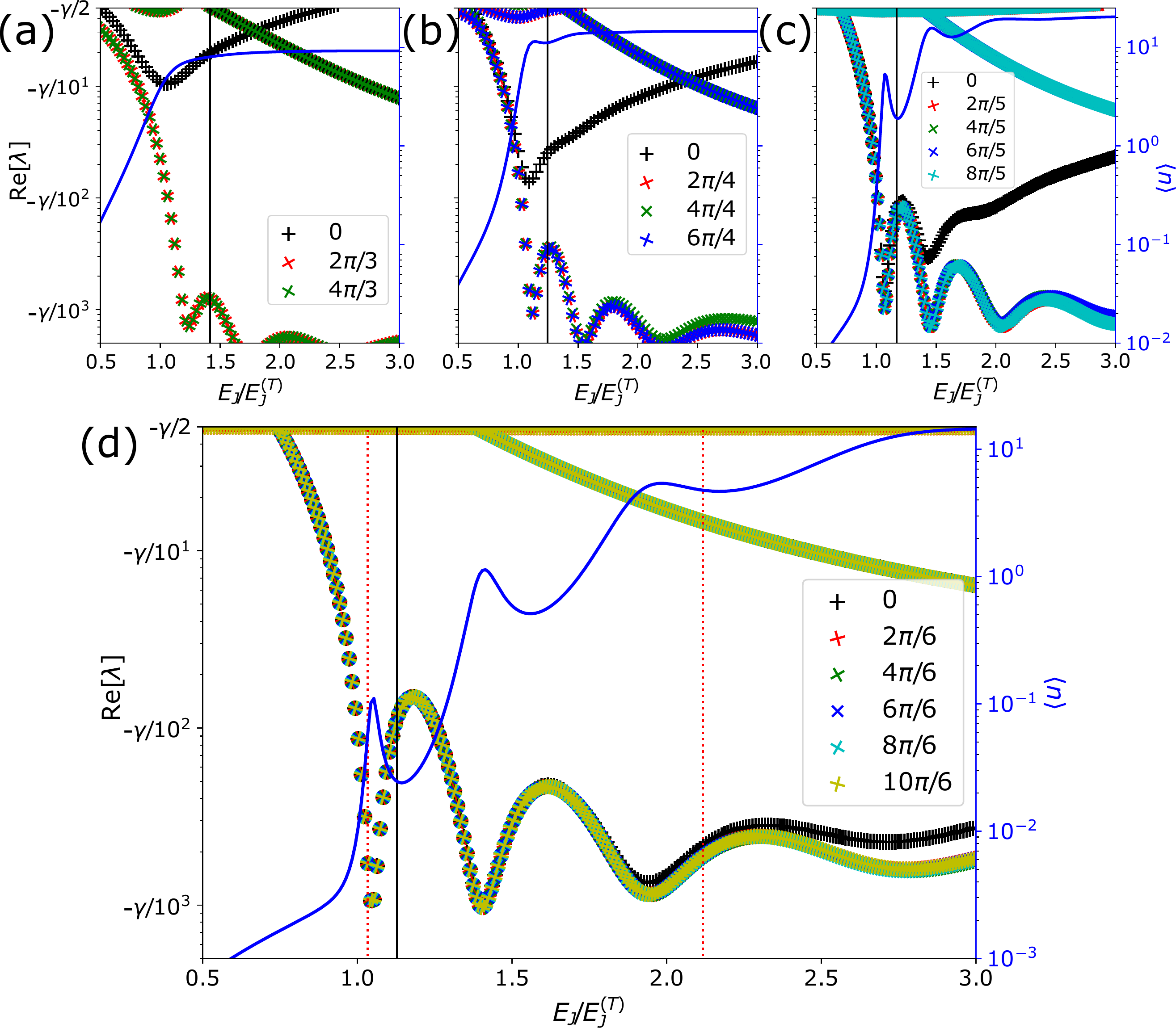}
\caption{Left axis: Lower part of the eigenvalue spectrum of the Liouvillian as a function of $\EJ$, with $\Delta=0.7$ (log scale). Data point colour indicates the eigenvalue of the corresponding eigenoperator under rotation. Black vertical line: location of the $2^{\text{nd}}$ bifurcation. Right axis: average steady-state photon occupation number (log scale). The behaviour for $p=3$, 4, 5 and 6 is shown in plots (a), (b, (c), and (d) respectively. The red-dotted vertical lines in (d) show two particular $\EJ$ values analysed in more detail in figures \ref{Complex_Wigners_1} and \ref{Complex_Wigners_2}.} 
\label{Spectrum}
\end{figure}
\subsection{Eigenvalues}
The behaviour of the Liouvillian eigenvalues with least negative (but non-zero) real parts is shown in figure \ref{Spectrum} as a function of the drive strength for $p=3,4,5,6$. These plots illustrate in particular how the emergence and subsequent evolution of slow time-scales in the system becomes connected with the oscillations in the average occupation number that develop for increasing $p$ values.

We start by considering the case where $p=3$ figure \ref{Spectrum}(a), where oscillations in $\langle n\rangle$ are not seen, and instead the log-scale makes clear that  $\langle n\rangle$ starts to saturate for $\EJ\sim\EJ^{(T)}$. There is also a clear descent in the real part of several eigenvalues towards zero as $\EJ^{(T)}$ is approached, with one eigenvalue (with rotational eigenvalue $k_n=0$) then rising up again\,\cite{Gosner_2020} whilst two others (with $k_n\neq 0$)  subsequently stay very small (i.e. with a real part that is very small in magnitude), albeit with a weak but noticeable oscillating component. This behaviour broadly matches what one would expect to see in the vicinity of a first-order dissipative transition in which symmetry breaking also occurs\,\cite{Minganti_2018}, and fits our expectations based on the underlying classical bifurcation. The features get sharper if one decreases $\Delta$, which increases $\langle n\rangle$ overall, taking us closer to the expected thermodynamic limit. Furthermore, a second set of eigenvalues start to drop towards zero signalling another dissipative transition around the onset of the second bifurcation. 

The behaviour becomes more complex as $p$ is increased, see figure \ref{Spectrum}(b,c,d) and fits less well with the standard behaviour associated with dissipative transition paradigms\footnote{The most natural explanation is that we are in fact moving further away from the thermodynamic limit  even though $\langle n\rangle$ actually increases with $p$. Interestingly, this in turn implies that the way in which this limit is approached varies with $p$.}. Although we still see a group of eigenvalues dropping towards zero around $\EJ\sim\EJ^{(T)}$, as $p$ increases the clear separation of the eigenvalues into those that stay very small and one that rapidly grows again (associated with a first order dissipative phase transition) breaks down. Instead, for $p=6$ the set of $6$ eigenvalues that drop towards zero stick together and then develop marked oscillations over a broad range of drive strengths. These oscillations match those seen in $\langle n\rangle$, revealing an apparent connection between the steady state behaviour and that of some of the slow dynamical time-scales in the problem. We will explore this connection in more detail in section \ref{sec:effective}, but for now we turn to look in more detail at the properties of the groups of eigenvalues that cluster together in figure \ref{Spectrum}(d) and properties of the corresponding eigenoperators. 

\subsection{Eigenoperators and Band Structure}

The eigenoperators of the Liouvillian reveal structures in phase space which closely resemble the structure of Bloch modes in a crystal. These Bloch-modes are similar to the states identified as phase space crystals\,\cite{Guo_2013,Guo_2016} for nonlinearly driven oscillators that are not subject to dissipation. However there are important differences as well. The modes discussed in the phase space crystal literature are simultaneous {\emph{eigenstates}} of the Hamiltonian and state-rotation operator, $\hat{H}_{\text{RWA}} \ket{\psi} = E \ket{\psi}$, $\hat{r} \ket{\psi} = \exp(\rmi k_n) \ket{\psi}$ \cite{Guo_2013}. In contrast, for the dissipative case, we are interested in operators that are simultaneous eigenoperators of the Liouvillian and operator-rotation,  $\Lagr \rho = \lambda \rho$, $\mathcal{R} \rho = \rme^{\rmi k_n} \rho$\,\cite{Gosner_2020}. Furthermore, whilst Bloch-like Hamiltonian eigenstates are distributed across the Hamiltonian maxima/minima\,\cite{Guo_2013}, the Liouvillian eigenoperators we investigate here are distributed across the classical fixed points of the system.

\begin{figure}[t]
\centering
\includegraphics[scale=0.35]{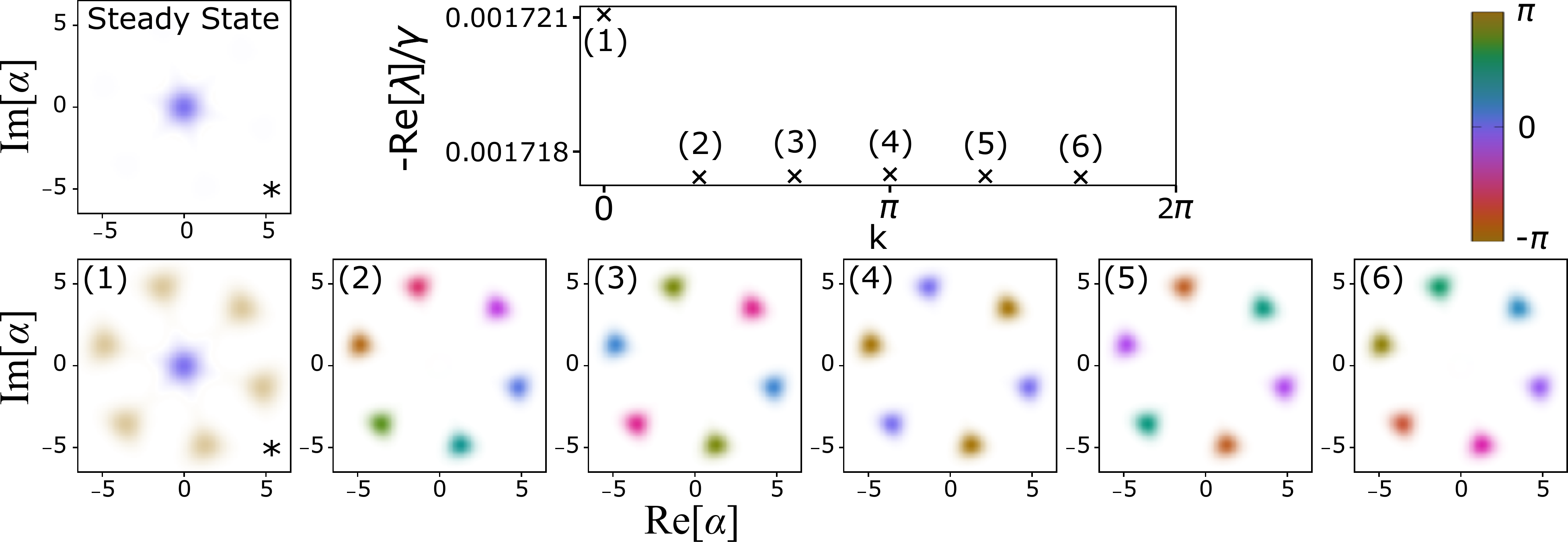}
\caption{Top right: Band diagram of 6 eigenvalues with smallest real parts and corresponding simultaneous eigenoperators of $\Lagr$ and $\mathcal{R}$. Here $\EJ \approx 1.03 \EJ^{(T)}$ [indicated by the first red dashed vertical line in figure \ref{Spectrum}d]. The steady-state and the eigenoperators labelled (1-6), following the numbering of eigenvalues in the band diagram, are shown as WDs. Colour intensity indicates the magnitude of complex WDs while hue indicates phase.  The steady state and (1) (indicated with an asterix) use the amplitude to the power of $1/2$ for the colour intensity to enhance the weaker features.}
\label{Complex_Wigners_1}
\end{figure}

In figure \ref{Complex_Wigners_1} we use the Wigner representation to depict the eigenoperators associated with the 7 eigenvalues with smallest (in magnitude) real parts. Note that transforming a Hermitian operator into phase space with the Wigner map produces a real valued field, so that a physical density operator corresponds to a real-valued Wigner function. However non-Hermitian operators give complex WDs in phase space, sometimes called non-diagonal Wigner functions\,\cite{QM_in_PS_book, Groenewold_1946}. In our case all operators with $k_n$ neither $0$ nor $\pi$ have complex WDs, as they must to have the correct (complex) eigenvalue under phase-space rotation. In each case the eigenoperators consist of parts centered at the classical fixed points, but with phases chosen to obey the rotational symmetry.

Taken together, the plots in figure \ref{Complex_Wigners_1} show that the group of points oscillating together in eigenvalue in figure \ref{Spectrum} correspond to a set of Bloch modes with differing $k_n$-value, but otherwise much alike. Their close lying eigenvalues suggest that coupling between BPs is very weak so that $k_n$ has very little effect on lifetime. Only the $k_n=0$ mode [mode (1) in figure \ref{Complex_Wigners_1}, shown as black crosses in figure \ref{Spectrum}] deviates significantly from the others: this difference arises because this mode is uniquely able to hybridise with the central DP (figure \ref{Complex_Wigners_1}). This splitting increases progressively with $\EJ$ (though at a rate that depends strongly on $p$), resulting eventually in a clear splitting-off of this eigenvalue from the others at higher $\EJ$ as can be seen in figure \ref{Spectrum}.

There is an equivalence between the clockwise and anti-clockwise directions in phase space for this Liouvillian, manifest as a degeneracy in eigenoperators of $\Lagr$ with equal and opposite rotational eigenvalues, $k_n$. By taking linear combinations inside these degenerate subspaces one could choose an alternative basis, for example one where all the eigenoperators of $\Lagr$ were Hermitian giving real WDs. However these eigenoperators would lack the rotational symmetry inherent to the problem, obscuring the important role this symmetry plays (the implications of basis choices are discussed in \ref{Appendix: Operator bases}).

In figure \ref{Complex_Wigners_2} we move to a higher value of $\EJ$, past the second bifurcation ($\EJ=2.12\EJ^{(T)}$). Here there are twice as many BPs in the steady-state, and their angular spacing alternates. Notice that in the Liouvillian eigenspectrum [figure \ref{Spectrum}(d)] at this $\EJ$ value there are {\emph{two}} groups of 6 almost degenerate eigenvalues with very small (negative) real values. The set that oscillates as a function of $\EJ$ (i.e.\ a continuation of those shown in figure \ref{Complex_Wigners_1} to larger drive values) and a new group which entered view shortly after the second bifurcation and which get progressively smaller (i.e.\ corresponding to longer lived excitations) as $\EJ$ is increased further. Plotting the WDs of the corresponding eigenoperators we see that the first set is analogous to the acoustic band of a diatomic crystal, while the other (shorter lived) set parallels the optical band \cite{stroscio_dutta_2001}. The key difference being in the former case the two `atoms' in each unit cell are in phase with one another while in the latter they are in anti-phase.
The emergence of the second group of nearly degenerate long-lived transients seen in figure \ref{Spectrum} is clearly associated with the formation of these optical modes.

\begin{figure}[t]
\centering
\includegraphics[scale=0.35]{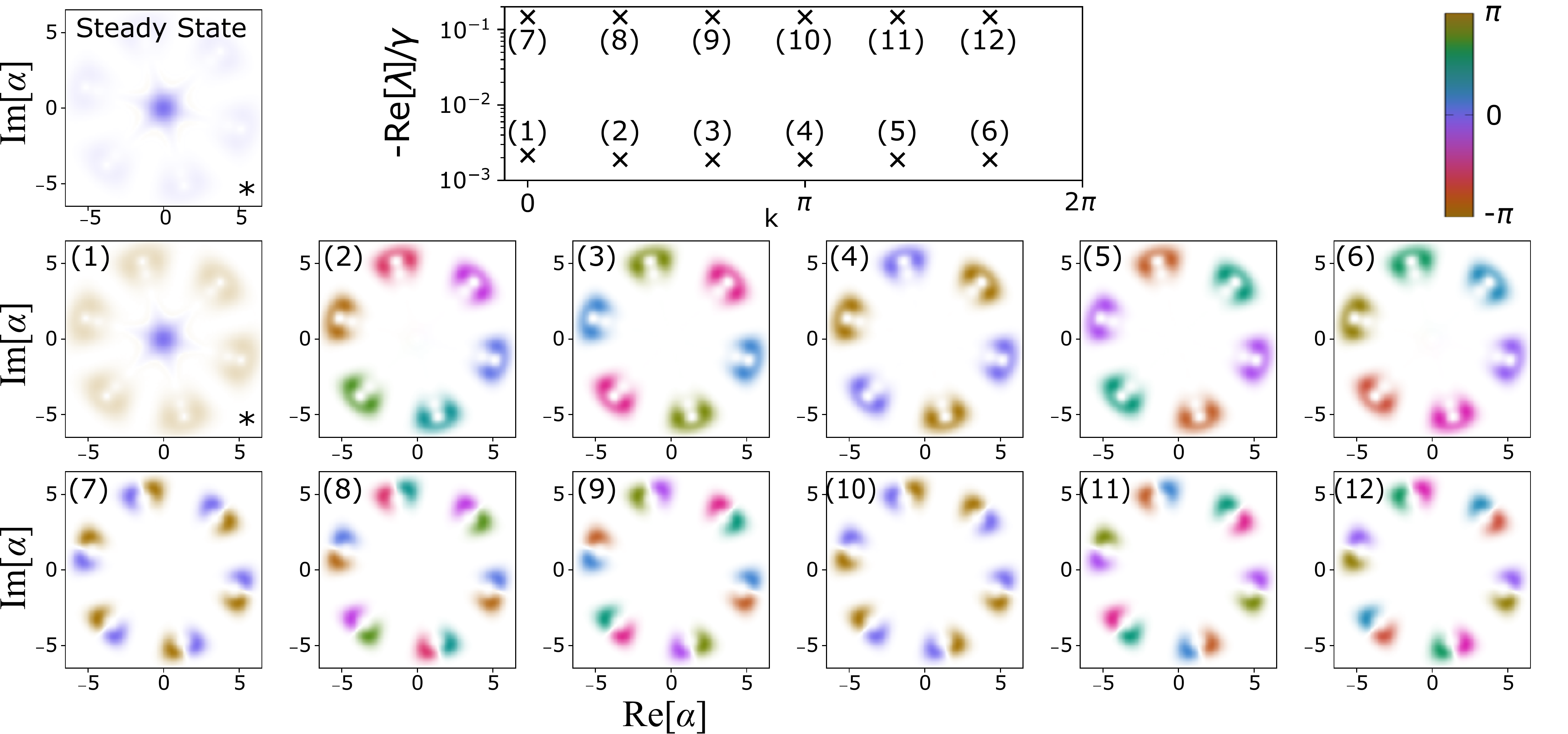}
\caption{Band diagram of 12 eigenvalues with smallest real parts (two lowest bands) and corresponding simultaneous eigenoperators of $\Lagr$ and $\mathcal{R}$. Here $\EJ \approx 2.12 \EJ^{(T)}$  [indicated by the second vertical dashed redline in figure \ref{Spectrum}d]  which is well above the second bifurcation and the eigenvalues are shown on a log-scale.}
\label{Complex_Wigners_2}
\end{figure}

\section{Effective Description}
\label{sec:effective}
The Liouvillian possesses an (in principle) infinite eigenvalue spectrum. However, as seen in figure \ref{Spectrum} a handful of these eigenvalues lead to decay rates of the associated eigenoperators that are much slower than $\gamma/2$ and well separated from the rest of the spectrum. This separation of timescales implies that the system relaxes rapidly towards a slow-dynamics subspace of much reduced dimension\,\cite{Macieszczak_2016,Katarzyna_2020}. In this section we show that the eigenoperators within this slow-subspace can be used to construct an effective description of the system's dynamics within phase space which in turn reveals that the oscillations in the occupation number are associated with oscillations in the rate describing motion of the system between the BPs and the DP. 


The long lived eigenoperators shown in figures \ref{Complex_Wigners_1} and \ref{Complex_Wigners_2} are concentrated in the vicinity of the stable classical fixed points. This implies the evolution of any initial WD will follow two stages, a rapid decay towards the nearest fixed point(s), and then a slow relaxation in which the weight of the WD is adjusted between the different fixed points to generate the correct balance of quasiprobability between these fixed points corresponding to the steady state.

\begin{figure}
\centering
\includegraphics[scale=0.6]{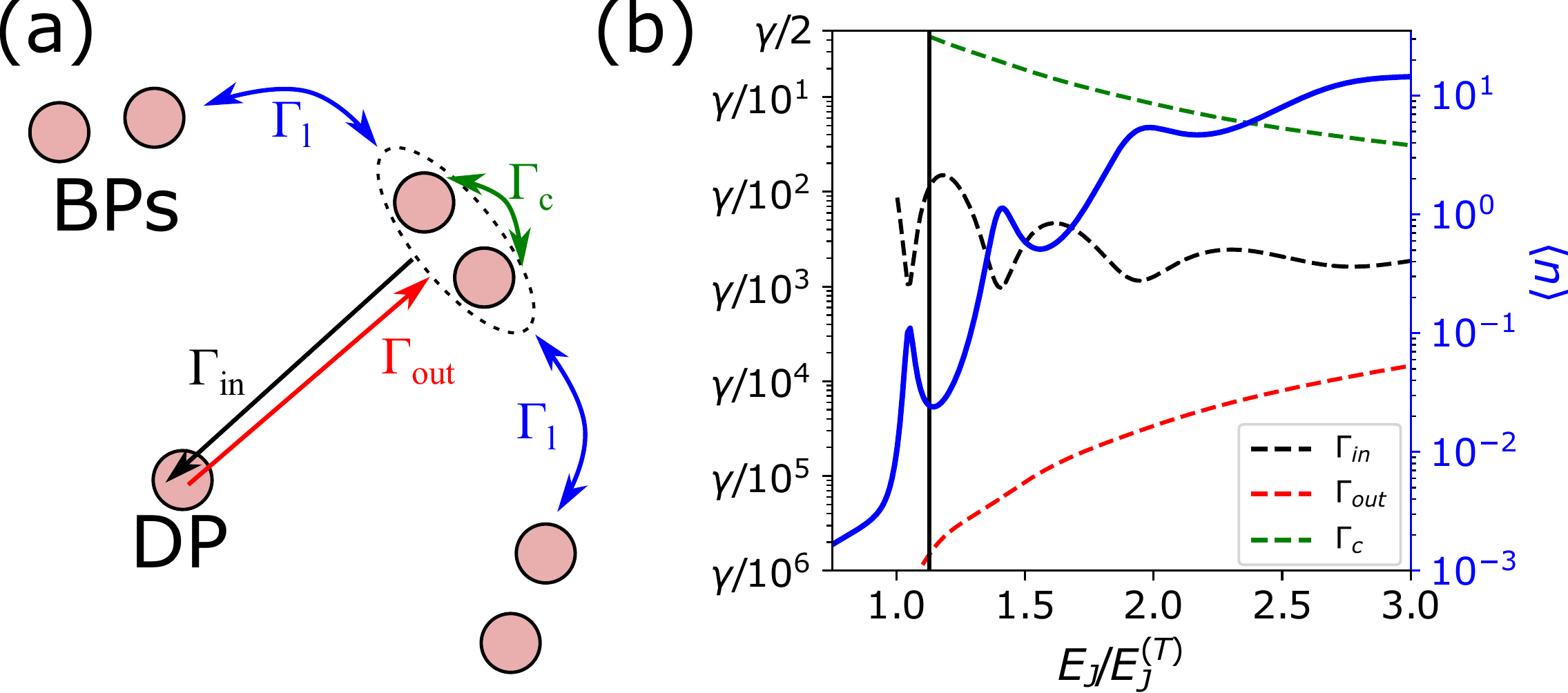}
\caption{(a) Cartoon indicating the meaning of the different rate constants in the effective model. (b) Values of the rate constants for a system with $p=6$ and $\Delta=0.7$. The rates $\Gamma_{\text{in}}$ and $\Gamma_{\text{out}}$ can only be obtained once the BPs have formed $\EJ>\EJ^{(T)}$, and $\Gamma_{\text{c}}$ above the second bifurcation (indicated by the vertical line). The inter-BP rate, $\Gamma_{\text{l}}$ falls below $\gamma/10^{6}$ over the entire parameter range and is not shown.}
\label{Interpoint_rates}
\end{figure}

This suggests that a simple effective model of the system's long term dynamics can be constructed. In this model the dynamics is encapsulated in a set of rate equations describing the rates at which the WD quasiprobability flows between the different states associated with each of the stable fixed points, assuming that each fixed point has rate constants connecting it to its nearest neighbours as illustrated in figure \ref{Interpoint_rates}(a). This implies four rates: one describing motion from the BPs to the DP, $\Gamma_{\text{in}}$, one for motion from the DP to the BPs, $\Gamma_{\text{out}}$, one for motion from one BP pair to a neighbouring pair, $\Gamma_{\text{l}}$, and finally (above the second bifurcation) a rate for motion from one member of a BP pair to its partner, $\Gamma_{\text{c}}$. This amounts to a reformulation of the problem from the reduced eigenbasis associated with the slow rates, to a `physical basis' associated with states localised in phase space.

The rates can be determined by mapping between the reduced eigenbasis and the physical basis. For example, the $k_n=0$ eigenoperator [see, e.g., figure \ref{Complex_Wigners_1}], consists of a central point with a positive amplitude surrounded by satellites of the opposite sign. This mode, like all others except the steady state, has a negative eigenvalue, so that time evolution results in uniform decay. 
Under time evolution WD quasiprobability is conserved locally \cite{Bauke_2011, Steuernagel_2013, Braasch_2019}, so the decay of the eigenmode must proceed by the positive and negative populations finding one another and annihilating. Using this to write down the rate equation in the physical basis, we can establish that the rate for transitions from a BP to the DP, $\Gamma_{\text{in}}$ and the reverse rate, $\Gamma_{\text{out}}$, must be related to the corresponding (real) eigenvalue as follows: $\lambda_{(1)} = - \Gamma_{\text{in}} - p\Gamma_{\text{out}}$. As shown in \ref{Appendix:Physical rates} similar arguments allow all of the decay rates in the simplified model to be determined in terms of the corresponding eigenmode decay rates. 

The resulting rates\footnote{Except $\Gamma_1$ which is extremely small throughout.} are shown in figure \ref{Interpoint_rates} for the case where $p=6$. We see that the rate of escape from the DP grows monotonically, and appears to be primarily responsible for the overall upward trend in $\langle n \rangle$ with $\EJ$. In contrast the rate inwards towards the DP oscillates and is clearly responsible for the peaks/oscillations in photon occupation. We can conclude that the peaks in photon number are driven by the suppression of a process, not the enhancement of one - a reduction in snakes, not an increase in ladders.

The rate describing coupling from one BP-pair to a neighbouring one ($\Gamma_{\text{l}}$) is significantly smaller than all the other rates (so much so that we have difficulty resolving it from zero) \footnote{Though, interestingly it can be significantly enhanced when dephasing to model fluctuations in the bias voltage is included, see \ref{Appendix:volt noise}}. This is exactly what we anticipated, based on the narrow band-width for the $k_n\neq 0$ states seen in figure \ref{Complex_Wigners_1}. Interestingly, the round-about process where amplitude leaves one BP pair to enter the DP ($\Gamma_{\text{in}}$), then leaves the DP to enter another BP pair ($\Gamma_{\text{out}}$) greatly dominates over direct movement from one pair to a neighbouring one. 

In addition to their impact on the equilibrium photon number these changing rates would have other detectable signatures. In equilibrium the system (for many parameter choices) occupies a mixture of the DP and the BPs. If one monitored the photon emission in these regimes one would see periods of high activity, where many photons are detected, corresponding to BP occupation. These bright periods would last an average duration of $1/\Gamma_{\text{in}}$ before giving way to dark periods of reduced activity which, in turn, would on average last $1/ \Gamma_{\text{out}}$ before the system flipped back to the bright phase. This flickering would imprint itself on, for example, a $g^{(2)}$ measurement \cite{Dambach_2015, Rolland_2019}. 

\section{Summary and Future Perspectives}
\label{sec:conclude}
We theoretically explored a superconducting system, comprised of a cavity coupled to a Josephson junction. We studied regimes where the dynamics lead to the creation/annihilation of up to $6$ photons at a time. The high level of rotational symmetry in phase space had several important effects. In the semi-classics it resulted in an `invincible' stable point at zero-amplitude, while in the quantum description it resulted in the Liouvillian eigenoperators having a Bloch mode structure in phase space, a structure one could describe as a dissipative phase space crystal.

An unusual quantum effect, where the expected number of photons in the cavity depended nonmonotonically on the drive/loss ratio was discovered and investigated. This behaviour arises from the full quantum coherent dynamics of the system and is not seen in the semi-classical limit described by a Fokker-Planck equation. We found that the oscillations in the occupation number were reflected in complex oscillatory changes in the rate at which the system can return to the vacuum from an excited state.


We hope that our work will stimulate detailed experimental studies of multi-photon resonances in superconducting circuits systems. The main features discussed throughout the paper should be experimentally accessible in a variety of currently available device architectures. For coplanar waveguide cavities like that in \cite{Chen_2014} the modest $\Delta$ means one needs a high, but attainable, drive-loss ratio ($E_J^{(T)} / (\hbar \gamma) \sim 3500$). High impedance resonators with $\Delta \approx 1$ \cite{Rolland_2019}, require much lower drive-loss ratios, 
$E_J^{(T)} / (\hbar \gamma) \approx 9.7$ for $p=6$. Resonances below $6$ will be more easily accessed. The key idealisations made by our model are assessed in appendices \ref{Appendix: RWA} (RWA) and \ref{Appendix:volt noise} (constant bias voltage), in both cases the results are not expected to change significantly when relaxing these assumptions. It would be interesting to investigate the extent to which oscillations in the cavity occupation number arise in set-ups that exploit flux rather than voltage bias\,\cite{Svensson_2017,Svensson_2018,Waltraut_2019,Brock_2020} where  clear evidence of several higher-order resonances (and corresponding periodic structures in the phase space) has already been seen\,\cite{Svensson_2017,Svensson_2018,Waltraut_2019}.

Finally, we note that from a theoretical perspective, future work is needed to provide a more intuitive understanding of how the oscillations in the photon occupation number arise. The fact that they only occur at relatively large values of the zero-point fluctuations where semi-classical approaches seem to break down makes the problem more challenging, but also more interesting.

\ack

We thank J. Ankerhold, M. Blencowe, S. Dambach and B. Kubala for very helpful conversations. The work was supported by a Leverhulme Trust Research Project Grant (RBG-2018-213).

\appendix

\section{Rotating Wave Approximation}
\label{Appendix: RWA}

Throughout this paper we have made use of the rotating wave approximation (RWA), including only stationary terms in the Hamiltonian within the rotating frame. In this Appendix we check the validity of this approach for multiphoton resonances by comparing with calculations carried out using the full time-dependent Hamiltonian.

For the non-RWA calculations,  we assume the initial state is the vacuum, and this is evolved forwards in blocks of 720 drive periods until it shows no significant change over a block. The expected photon number and photon number histograms are averaged over a single period.

\begin{figure}[h!]
\centering
\includegraphics[scale=0.45]{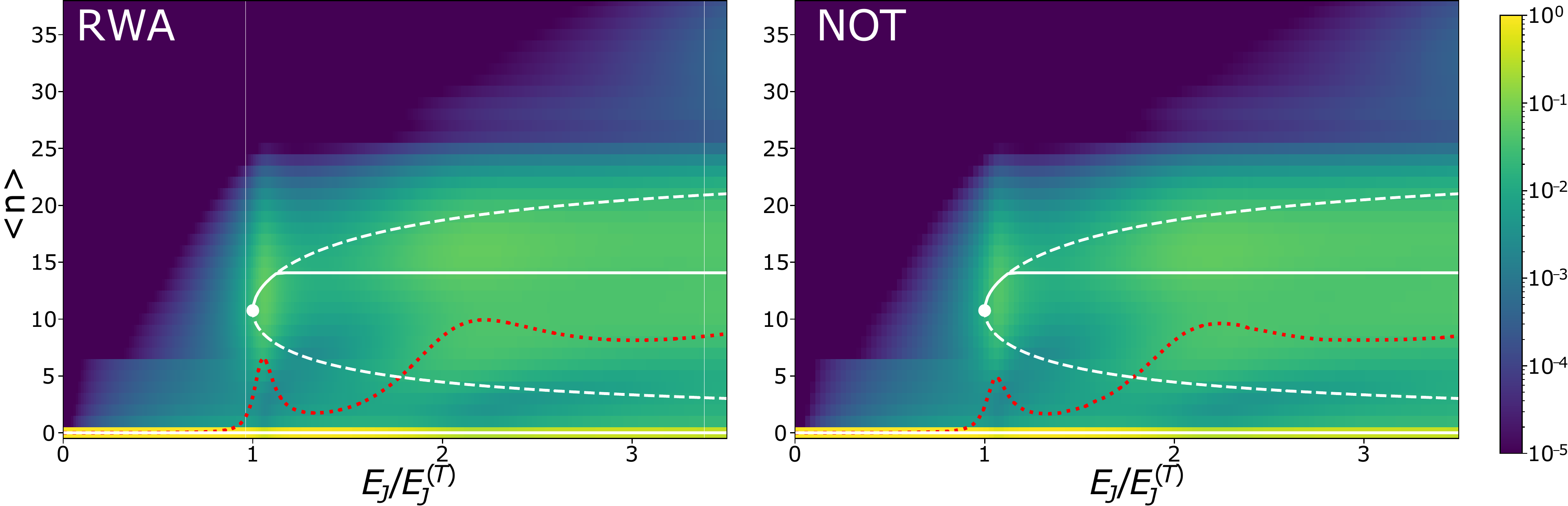}
\caption{Photon numbers in the steady state as a function of $\EJ$ for $\Delta = 1$, $p=6$ with $\omega_J=6\omega_0$ and $\omega_0 /\gamma= 1\times10^3$ with (left panel) and without (right panel) the RWA. The colour (on a log-scale) indicates $P_n=\langle n|\rho|n\rangle$,  also shown are the average photon number (red-dashed line), together with the squared amplitudes of the stable (solid white lines) and unstable (dashed white lines) fixed points.}
\label{RWA_fig}
\end{figure}

RWA and non-RWA results for $p=6$ are compared in figure \ref{RWA_fig}. It can be seen that even for the relatively high quantum fluctuation strength chosen ($\Delta=1$) there is quite close agreement. The main difference is that the sharp peak in excitation number, while still clearly present, is somewhat less pronounced in the non-RWA solution.

We note, however, that the RWA eventually fails at very low values of $\EJ$ (invisible on the scale of figure \ref{RWA_fig}). This is because very far below threshold the resonant $p$-photon process is suppressed practically to zero, so that the (very far) off-resonant 1-photon process neglected in the RWA is no longer completely negligible in comparison \cite{Morley_2019}.

\section{Voltage Noise}
\label{Appendix:volt noise}

In the main text we assume that the bias voltage, $V$, is fixed. This is an idealisation of the situation found in experiments where additional impedances in the circuit give rise to fluctuations in the voltage seen by the JJ-cavity system\,\cite{Gramich_2013,Souquet_2016,Wang_2017}. In this Appendix we use a simple approximate description\,\cite{Souquet_2016} which maps the voltage fluctuations onto a fluctuating cavity frequency which in turn leads to photon-number dephasing.

\begin{figure}[h!]
\centering
\includegraphics[scale=0.55]{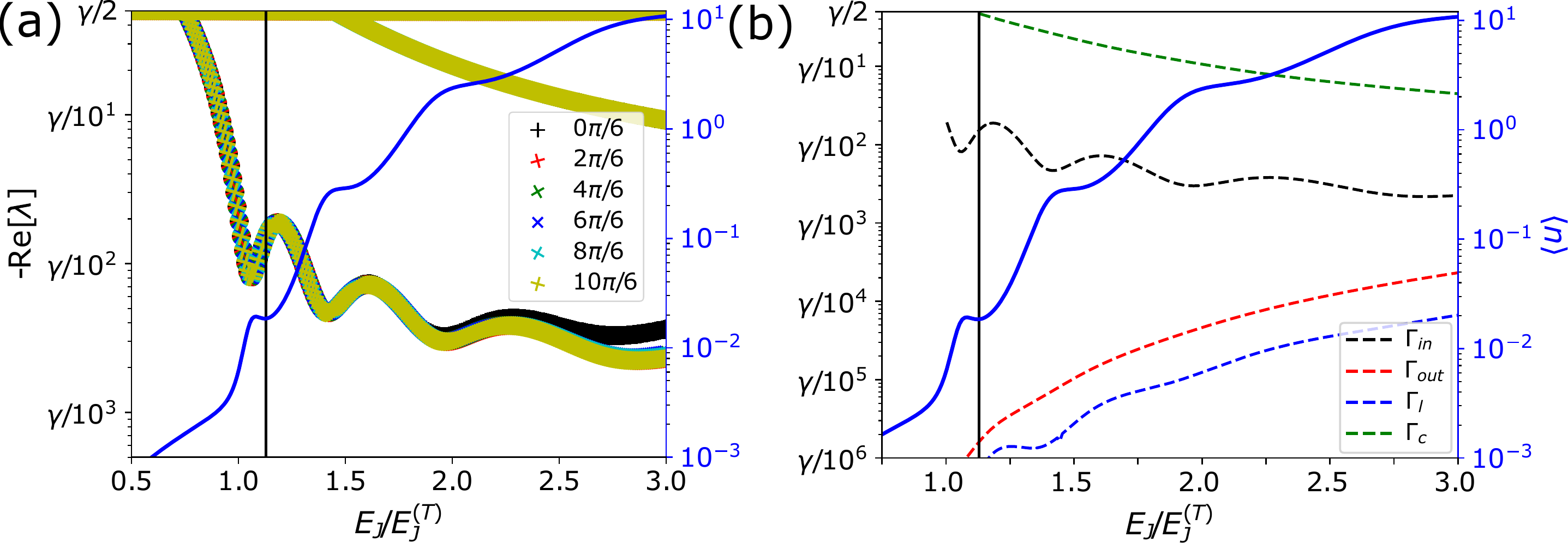}
\caption{Average occupation number (blue) shown together with (a) the lowest part of the Liouvillian eigenspectrum together with (b) the extracted rates for $\gamma_{\text{VN}} / \gamma = 0.05$, $\Delta = 0.7$, $p=6$. The corresponding plots obtained without including voltage noise are shown in figures \ref{Spectrum}(d) and \ref{Interpoint_rates}(b).}
\label{Volt_noise}
\end{figure}

Assuming that the dephasing due to the voltage fluctuations is very weak\,\cite{Souquet_2016} leads 
to the modified Liouvillian:
\begin{equation}
\Lagr_{\text{VN}} = \Lagr + \frac{\gamma_{\text{VN}} }{2} ( 2 \hat{n}_p \rho \hat{n}_p  - \hat{n}_p^2 \rho - \rho \hat{n}_p^2 ),
\label{Modified_Master_Equation}
\end{equation}
with $\hat{n}_p = \cre \des / p$. 
Results obtained using this Liouvillian are shown in figure \ref{Volt_noise} for the case where $\gamma_{\text{VN}} / \gamma = 0.05$, chosen to be similar to that in some experiments \cite{Gramich_2013}. Notice that the sharp features in $\langle n \rangle$ have been suppressed by the voltage noise, though the peak structure is still clearly present.  Figure \ref{Volt_noise} also shows the corresponding eigenspectrum of the modified Liouvillian. 

Figure \ref{Volt_noise}b shows the impact of the voltage noise on the rates for our simple effective model (see sec.\ \ref{sec:effective}). The coupling from one of the BP pair to its neighbours, $\Gamma_{\text{l}}$ is much higher. While it remains the slowest rate it is now large enough to fit a value to it. Recall that, without voltage noise, this rate is so small that we could not reliably resolve it from zero. The increase in this rate is perhaps not surprising as the main effect of photon-number dephasing is essentially angular diffusion.


\section{Wigner transform}
\label{Appendix:Wigner transform}

This Appendix describes the quantum Fokker-Planck equation used for a semi-classical understanding of the system. 

The quantum Fokker-Planck equation describing the evolution in phase space is obtained from the master equation (equation \ref{Master_equation}) with the following substitutions \cite{QM_in_PS_book, Groenewold_1946}:
\begin{equation}
\begin{split}
\rho  &\rightarrow W(\alpha, \alpha^*) \\
\des, \, \cre & \rightarrow \alpha, \, \alpha^* \\
( \hat{A}(\des, \cre) )_{S} &\rightarrow \mathcal{A}(\alpha, \alpha^*) \\
\hat{A}\hat{B} &\rightarrow \mathcal{A}\star\mathcal{B},
\label{Phase_Map}
\end{split}
\end{equation}
hence an appropriately symmetrised operator (indicated by the subscript $S$) becomes the corresponding function of the complex amplitudes.  The star product is defined as:
\begin{equation}
\mathcal{A} \star \mathcal{B} = \mathcal{A}  \exp\left( \frac{1}{2} \left( \overleftarrow{\frac{\partial}{\partial \alpha}}\overrightarrow{\frac{\partial}{\partial \alpha^*}} - \overleftarrow{\frac{\partial}{\partial \alpha^*}}\overrightarrow{\frac{\partial}{\partial \alpha}} \right) \right) \mathcal{B},
\end{equation}
with the arrows indicating the direction in which the derivatives act and the exponentiation of the derivatives is understood in terms of the Taylor series.

It is worth explaining the nature of these replacements. The density operator (normally represented as a matrix) is replaced with the WD, which contains identical information in an alternative format \cite{Carmichael_book}. The new expression of the Hamiltonian, $\mathcal{H}$, is given by the Wigner transform of the Hamiltonian operator. This is gained by first arranging the Hamiltonian such that its operators are in symmetric order (indicated by the subscript $_S$ in equation \ref{Phase_Map}). Then the appropriate Hamiltonian for the Fokker-Planck model is found by a simple substitution of $\des$ and $\cre$ with $\alpha$ and $\alpha^*$ \cite{Carmichael_book}.

For our Hamiltonian moving between normal order (equation \ref{rwaham}) and symmetrical order changes the expression very little. The only change is that the factor of $\exp(-\Delta^2/2)$ present in the normally ordered version disappears. The easiest way of confirming this is to show that the full, time-dependent, Hamiltonian in equation (\ref{time_Hamiltonian}) (which is already expressed in symmetrical order) gains this factor when re-arranged into normal order using the Baker-Hausdorff rule \cite{mandel_wolf_book}. Although this relationship can also be confirmed in the rotating frame using the expressions in \cite{Carmichael_book}.

These substitutions produce the following equation of motion for the WD
\begin{equation}
\begin{split}
\frac{ \partial W}{\partial t} =& \frac{2 \mathcal{H}}{\hbar} \sin \left( \frac{1}{2\rmi} \left( \overleftarrow{\frac{\partial}{\partial \alpha}} \overrightarrow{\frac{\partial}{\partial \alpha^*}} -
\overleftarrow{\frac{\partial}{\partial \alpha^*}} \overrightarrow{\frac{\partial}{\partial \alpha}}\right) \right) W  \\
& + \frac{\gamma}{2} \left[ \frac{\partial }{\partial \alpha} \alpha + \frac{\partial }{\partial \alpha^*} \alpha^*  + \frac{\partial^2 }{\partial \alpha \partial \alpha^*} \right] W,\\
\end{split}
\label{Derivative_series_equation}
\end{equation}
with $\mathcal{H} = \hbar \delta A^2 - \EJ J_p(2\Delta A) \cos(p\theta + p\pi/2)$. This Hamiltonian is identical to the classical rotating wave Hamiltonians given in \cite{Guo_2016, Armour_2017} to describe equivalent systems.

The terms related to first derivatives in (\ref{Derivative_series_equation}) (both from the Hamiltonian and loss) are \emph{drift terms}. 
The coefficients of these terms give the classical trajectories. 

These trajectories are almost identical to the trajectories predicted by the coherent state ansatz, The difference is that the coherent state model posses an extra factor $\rme^{-\Delta^2/2}$ appearing next to the terms that derive from the Hamiltonian. The reason for this difference is that in the Fokker-Planck model the drift evaluated at some specific point ($\alpha$, $\alpha^*$) reflects the drift motion for the part of the WD at precisely that location. In contrast the coherent state ansatz equation of motion evaluated at ($\alpha$, $\alpha^*$) reflects the motion of a coherent state (Gaussian WD) with centre at that location.


The drift terms pull the WD towards the attractors (stable fixed points). However, they compete with the second derivative, \emph{diffusion}, terms which encourage it to spread out. The third-and-higher derivatives are quantum terms, when they are neglected (an approach known as the truncated Wigner approximation) we obtain a Fokker-Planck equation as discussed in the main text.

In the phase-space description the classical approximation amounts to taking only the first term in the Taylor series of the $\sin$ function in (\ref{Derivative_series_equation})\cite{QM_in_PS_book}. In our case the parameter $\Delta$ sets the radial scale of $\mathcal{H}$ relative to fundamental scale, $\hbar$. Thus the parameter $\Delta$ can be thought of as the system's `quantumness' \cite{Guo_2016}. Similarly $p$ sets the angular scale, so quantum effects are stronger at higher $p$ (each differentiation of $\mathcal{H}$ with respect to angle will bring out another factor of $p$).

\begin{figure}[h!]
\centering
\includegraphics[scale=0.8]{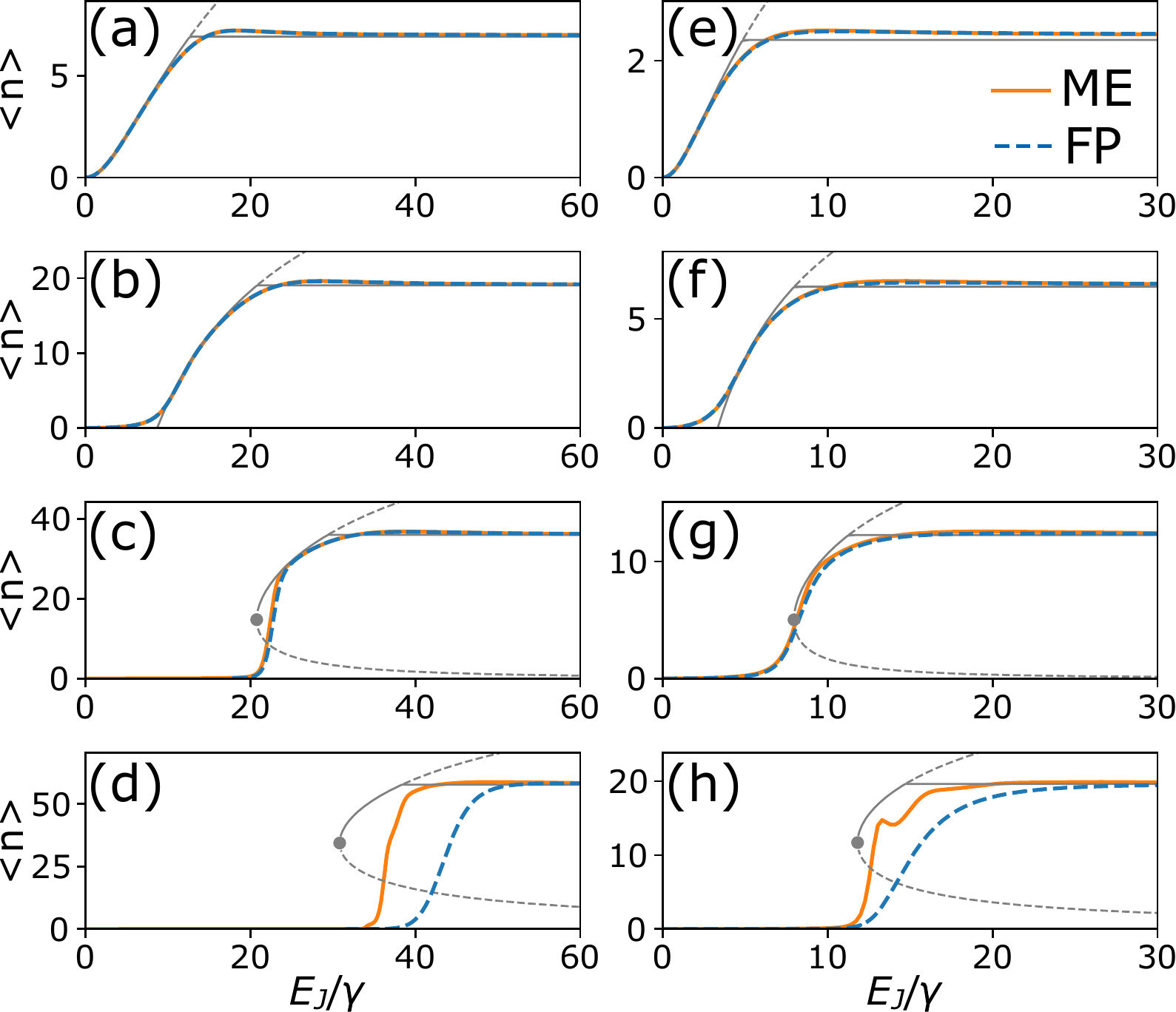}
\caption{Expected photon numbers calculated using the Fokker-Planck equation (equation \ref{Derivative_series_equation}) in blue and master equation (\ref{Master_equation}) in orange. The solid/dashed grey curves show the stable/unstable fixed points (obtained using the coherent state ansatz). (a-d) $\Delta =0.35$, $p=1$-4. (e-h) $\Delta = 0.6$, $p=1$-4.}
\label{FP_vs_QT}
\end{figure}

The Fokker-Planck (FP) equation for the WD, including only 1st and 2nd derivatives, was solved numerically using the FIPY package \cite{FIPY}. Expected photon numbers are plotted in figure \ref{FP_vs_QT}. Lacking higher derivatives these Fokker-Planck solutions are essentially classical, but with added noise. For the lower choices of $p$ these solutions match up almost perfectly with numerical solutions of the full master equation as can be seen in panels (a, b, e, f) of figure \ref{FP_vs_QT}.

For $p=3$ we see some deviation, with the FP solutions remaining near the DP slightly longer. This trend grows more pronounced at $p=4$. Importantly the Fokker-Planck solutions lack the peak structures, confirming the quantum origin of these features.

\section{Stability and Bifurcations}
\label{Appendix: Bifurcations}

In this Appendix we obtain analytic expressions for the threshold $\EJ$ values at the saddle-node bifurcations which mark the first appearance of fixed points away from the origin (for $p>2$). These bifurcations occur when both the drift terms in the Fokker-Planck equation and the determinant of the corresponding Jacobian are zero.

The drift terms of equation \ref{Derivative_series_equation} are conveniently expressed in cylindrical coordinates using $\overleftarrow{\partial_{\alpha}} \overrightarrow{\partial_{\alpha^*}} - \overleftarrow{\partial_{\alpha^*}} \overrightarrow{\partial_{\alpha}} =  \overleftarrow{\partial_{A}} (\rmi /  2A) \overrightarrow{\partial_{\theta}} - \overleftarrow{ \partial_{\theta}} (\rmi /  2A) \overrightarrow{\partial_{A}}$:
\begin{equation}
\vec{D}  =  \frac{1}{2}\begin{pmatrix} 


(-1/A\hbar) \partial_{\theta} \mathcal{H} + \gamma A \\
(1/\hbar) \partial_A \mathcal{H}
\end{pmatrix}.
\label{Drift}
\end{equation}
$\vec{D}$ is defined so that $\dot{W} = \nabla \cdot (\vec{D} W) + O(\partial^{2+}W)$, recalling that in cylindrical coordinates $\nabla = \begin{pmatrix} (1/A) \partial_A A & (1/A) \partial_{\theta} \end{pmatrix}$. The first (second) element is the radial (angular) part.

The Jacobian related to this drift is:
\begin{equation}
\overline{J} = \begin{pmatrix} \partial_A & \partial_{\theta} \end{pmatrix} \otimes \vec{D} = \frac{1}{2} \begin{pmatrix} 
(-1/A\hbar)  \partial_A \partial_{\theta} \mathcal{H} + (1 / A^2\hbar) \partial_{\theta} \mathcal{H} + \gamma &
(-1/A\hbar) \partial_{\theta}^2 \mathcal{H} \\
(1/\hbar) \partial_{A}^2 \mathcal{H} &
(1/\hbar) \partial_A \partial_{\theta} \mathcal{H}
\end{pmatrix}.
\label{Jacobian}
\end{equation}

Choosing zero detuning ($\delta=0$) implies that for low amplitudes the first fixed points to emerge away from the origin (as $\EJ$ is increased) all lie along specific angles\,\cite{Armour_2013}: $\sin(p(\theta + \pi/2)) = 1$. 
For these angles most of the derivatives in equation (\ref{Jacobian}) vanish, only those involving exactly one derivative with respect to $\theta$ remain. Setting both the drift and the determinant of the Jacobian to zero we find that the fixed point amplitude at the bifurcation satisfies:
\begin{equation}
\Delta A_r J^{'}_{p}(2\Delta A_r) = J_{p}(2\Delta A_r),
\label{A_crit}
\end{equation}
and the threshold value of $\EJ$ at which the bifurcation occurs is
\begin{equation}
\EJ^{(T)}  \hspace{0.5pc} = \frac{\gamma A_r^2 \hbar}{p J_p(2\Delta A_r)}.
\label{EJ_crit}
\end{equation}
This gives the threshold $\EJ$ at which the FP drift terms show a bifurcation. Notice, however, that the $\EJ$ at which the corresponding bifurcation occurs using the coherent state anstaz (see Sec.\ \ref{sec:fixedpoints}) drifts is slightly higher, by the factor $\rme^{\Delta^2/2}$. It is these values (including the extra factor) that are plotted by the white circles in figure \ref{photon_numbers}. Interestingly, expressions of the form of (\ref{A_crit}) apparently arise quite often \cite{LANDAU_1999}.

\section{Operator Bases}
\label{Appendix: Operator bases}

QuTiP was used to find the eigenoperators and eigenvalues of the Liouvillian \cite{QUTIP}. The symmetry under $\mathcal{R}$ of these eigenoperators was checked and it was found that, for any eigenoperator with a unique $\Lagr$-eigenvalue $\lambda_i$, the symmetry was obeyed with rotation eigenvalue either $1$ or $-1$ ($k_n =0$ or $\pi$, the two unpaired modes with no direction of circulation around phase space).

The remainder of the numerically discovered eigenoperators occur in sets that are degenerate with respect to the Liouvillian (within numerical precision), which means that any linear combination of eigenoperators from a given set is an equally valid eigenoperator. Thus the operators found numerically are valid, but in the wrong basis to have the rotational symmetry we expect. In order to find a basis of operators that possess this symmetry we first use Gram-Schmidt orthonormalization to find orthonormal operators that span the same space as each of these degenerate sets. Specifically we isolate a set of operators that are $\Lagr$-degenerate to within our numerical precision, $\{ \rho^{\text{num}}\}$, from this we generate the elements of a new basis $\{ \rho^{\text{GS}}\}$ by:
\begin{equation}
\rho^{\text{GS}}_n = \rho^{\text{num}}_n -  \sum_{x=0}^{n-1}   \rho^{\text{GS}}_x \frac{ \Tr[ (\rho^{\text{num}}_n)^{\dag}   \rho^{\text{GS}}_x ] }{ \Tr[ (\rho^{\text{GS}}_x)^{\dag}   \rho^{\text{GS}}_x ]}.
\end{equation}

These basis operators will still not have the required symmetries, but they will be orthogonal allowing us to proceed to the next step, where the rotation super-operator $\mathcal{R}$ is determined as a matrix in this GS basis. Writing some linear combination of operators $a \rho^{\text{GS}}_1 + b \rho^{\text{GS}}_2 + c \rho^{\text{GS}}_3 ..$ as a vector $\vec{B} = \begin{pmatrix} a & b & c & \dots \end{pmatrix}$ we see:
\begin{equation}
\vec{A} = \mathcal{R} \vec{B}    \hspace{2pc} \rightarrow \hspace{1pc} \vec{A} = \begin{pmatrix} 
\Tr[ (\rho^{\text{GS}}_1)^{\dag}   \mathcal{R}\rho^{\text{GS}}_1 ] & \Tr[ (\rho^{\text{GS}}_1)^{\dag}   \mathcal{R}\rho^{\text{GS}}_2 ] & \dots  \\
\Tr[ (\rho^{\text{GS}}_2)^{\dag}   \mathcal{R}\rho^{\text{GS}}_1 ] & \Tr[ (\rho^{\text{GS}}_2)^{\dag}   \mathcal{R}\rho^{\text{GS}}_2 ] & \dots   \\
\vdots                                                                   & \vdots                                                                   & \ddots  \\
\end{pmatrix} \vec{B}
\end{equation}
The eigenvalues of this matrix representation of $\mathcal{R}$ are a subset of the eigenvalues of $\mathcal{R}$ (specifically the subset that is degenerate with respect to $\Lagr$). The eigenvectors of this matrix represent the symmetry basis-operators in terms of the Gram-Schmidt basis. Thus we have converted the numerical eigenoperators into a new basis where each eigenoperator of $\Lagr$ is also an eigenoperator of $\mathcal{R}$. These are the operators whose WDs are depicted in figures \ref{Complex_Wigners_1} and \ref{Complex_Wigners_2}.

\section{Physical rates}
\label{Appendix:Physical rates}

This Appendix presents the equations linking the physical rates in our effective description of the system dynamics to the Liouvillian eigenvalues. Initially specialising to the regime between the two bifurcations, where there are only $p$ BPs we can express the point-to-point coupling with two differential equations:
\begin{equation}
\begin{split}
\dot{A}_m &= \Gamma_{\text{l}}( - 2 A_m + A_{m+1} +  A_{m-1} ) - \Gamma_{\text{in}} A_m + \Gamma_{\text{out}} D \\
\dot{D} &= - p \Gamma_{\text{out}} D + \Gamma_{\text{in}} \left(\sum_m A_m \right),
\end{split}
\label{Rates}
\end{equation}
with $A_m$ the amplitude at the $m^{\text{th}}$ BP and $D$ that at the DP.

We now find the initial choices of $D(0)$, $A_{m}(0)$ that result in all these quantities decaying together as $\rme^{-\lambda t}$ in response to these differential equations. They can be found by setting $\lambda = \dot{A}_m / A_m = \dot{D} / D$ and re-arranging.

First assume that all $A_{m}$s are the same and notice that $D(0) = 1$, $A_m(0) = -1/p$ satisfies this relation. This is the $k=0$ eigenoperator and has decay rate $\lambda = -p\Gamma_{\text{out}} - \Gamma_{\text{in}}$.

Assuming that $D(0)=0$ leads to $p$ solutions, one for each way of choosing $A_{m+1} = \exp(\rmi k) A_{m}$ such that $\sum_m A_m = 0$. These have decay times $\lambda(k) = \Gamma_{\text{l}} (2 \cos(k) - 2) - \Gamma_{\text{in}}$. For these modes $\Gamma_{\text{l}}$ produces band dispersion, although for our system we find $\Gamma_{\text{l}}$ to be very small. These solutions correspond to the eigenoperators depicted in figure \ref{Complex_Wigners_1}.

Finally we generalise to the case after the second bifurcation, where each BP splits into two. To retain the previous rate definitions we now define $A_m$ to be the amplitude across a BP pair. Assuming the pair to be in phase leaves all solutions completely unchanged from the previous (pre-bifurcation) results.

There are now new solutions, corresponding to assuming the two BPs in the pair have opposite phases. Here two things are different. First a new internal decay rate exists, where amplitude from each of the BPs forming the pair annihilates with that from the other. Second, the interactions with the neighbouring pairs acquire a "$-$" sign, as each BP has an extra $-1$ factor relative to its neighbours. The new set of solutions have decay rates $\lambda(k) = \Gamma_{\text{l}} (-2 \cos(k) - 2) - \Gamma_{\text{in}} - 2 \Gamma_{\text{c}}$, (these new ones include $k=0$).

These relations are exploited in reverse to fit $\Gamma_{\text{in}}$, $\Gamma_{\text{out}}$, $\Gamma_{\text{l}}$ and $\Gamma_{\text{c}}$ to the known eigenvalues of the Liouvillian.

\section*{References}
\bibliographystyle{unsrt}
\bibliography{bibliography}

\providecommand{\newblock}{}
\begin{thebibliography}{10}
\expandafter\ifx\csname url\endcsname\relax
  \def\url#1{{\tt #1}}\fi
\expandafter\ifx\csname urlprefix\endcsname\relax\def\urlprefix{URL }\fi
\providecommand{\eprint}[2][]{\url{#2}}

\bibitem{Mundhada_2019}
Mundhada S, Grimm A, Venkatraman J, Minev Z, Touzard S, Frattini N, Sivak V,
  Sliwa K, Reinhold P, Shankar S, Mirrahimi M and Devoret M 2019 {\em Phys.
  Rev. Applied\/} {\bf 12}(5) 054051

\bibitem{Gottesman_2001}
Gottesman D, Kitaev A and Preskill J 2001 {\em Phys. Rev. A\/} {\bf 64}(1)
  012310

\bibitem{Braunstein_1987}
Braunstein S~L and McLachlan R~I 1987 {\em Phys. Rev. A\/} {\bf 35}(4)
  1659--1667

\bibitem{Chang_2020}
Chang C~W~S, Sab\'{\i}n C, Forn-D\'{\i}az P, Quijandr\'{\i}a F, Vadiraj A~M,
  Nsanzineza I, Johansson G and Wilson C~M 2020 {\em Phys. Rev. X\/} {\bf
  10}(1) 011011

\bibitem{Guo_2013}
Guo L, Marthaler M and Sch\"on G 2013 {\em Phys. Rev. Lett.\/} {\bf 111}(20)
  205303

\bibitem{Guo_2020}
Guo L and Liang P 2020 {\em New Journal of Physics\/} {\bf 22} 075003

\bibitem{Zhang_2017}
Zhang Y, Gosner J, Girvin S~M, Ankerhold J and Dykman M~I 2017 {\em Phys. Rev.
  A\/} {\bf 96}(5) 052124

\bibitem{Zhang_2019}
Zhang Y and Dykman M~I 2019 {\em Phys. Rev. E\/} {\bf 100}(5) 052148

\bibitem{Lorch_2019}
L\"orch N, Zhang Y, Bruder C and Dykman M~I 2019 {\em Phys. Rev. Research\/}
  {\bf 1}(2) 023023

\bibitem{Gosner_2020}
Gosner J, Kubala B and Ankerhold J 2020 {\em Phys. Rev. B\/} {\bf 101}(5)
  054501

\bibitem{Tadokoro_2020}
Tadokoro Y, Tanaka H and Dykman M~I 2020 {\em Scientific Reports\/} {\bf 10}(1)
  10413

\bibitem{Guo_2016}
Guo L and Marthaler M 2016 {\em New Journal of Physics\/} {\bf 18} 023006

\bibitem{Hofheinz_2011}
Hofheinz M, Portier F, Baudouin Q, Joyez P, Vion D, Bertet P, Roche P and
  Esteve D 2011 {\em Phys. Rev. Lett.\/} {\bf 106}(21) 217005

\bibitem{Armour_2013}
Armour A~D, Blencowe M~P, Brahimi E and Rimberg A~J 2013 {\em Phys. Rev.
  Lett.\/} {\bf 111}(24) 247001

\bibitem{Gramich_2013}
Gramich V, Kubala B, Rohrer S and Ankerhold J 2013 {\em Phys. Rev. Lett.\/}
  {\bf 111}(24) 247002

\bibitem{Chen_2014}
Chen F, Li J, Armour A~D, Brahimi E, Stettenheim J, Sirois A~J, Simmonds R~W,
  Blencowe M~P and Rimberg A~J 2014 {\em Phys. Rev. B\/} {\bf 90}(2) 020506

\bibitem{Leppakangas_2015}
Lepp\"akangas J, Fogelstr\"om M, Grimm A, Hofheinz M, Marthaler M and Johansson
  G 2015 {\em Phys. Rev. Lett.\/} {\bf 115}(2) 027004

\bibitem{Cassidy_2017}
Cassidy M~C, Bruno A, Rubbert S, Irfan M, Kammhuber J, Schouten R~N, Akhmerov
  A~R and Kouwenhoven L~P 2017 {\em Science\/} {\bf 355} 939--942

\bibitem{Rolland_2019}
Rolland C, Peugeot A, Dambach S, Westig M, Kubala B, Mukharsky Y, Altimiras C,
  le~Sueur H, Joyez P, Vion D, Roche P, Esteve D, Ankerhold J and Portier F
  2019 {\em Phys. Rev. Lett.\/} {\bf 122}(18) 186804

\bibitem{Svensson_2017}
Svensson I~M, Bengtsson A, Krantz P, Bylander J, Shumeiko V and Delsing P 2017
  {\em Phys. Rev. B\/} {\bf 96}(17) 174503

\bibitem{Waltraut_2019}
Wustmann W and Shumeiko V 2019 {\em Low Temperature Physics\/} {\bf 45}
  848--869

\bibitem{Meister_2015}
Meister S, Mecklenburg M, Gramich V, Stockburger J~T, Ankerhold J and Kubala B
  2015 {\em Phys. Rev. B\/} {\bf 92}(17) 174532

\bibitem{Kubala_2015}
Kubala B, Gramich V and Ankerhold J 2015 {\em Physica Scripta\/} {\bf T165}
  014029

\bibitem{Wang_2017}
Wang H, Blencowe M~P, Armour A~D and Rimberg A~J 2017 {\em Phys. Rev. B\/} {\bf
  96}(10) 104503

\bibitem{Minganti_2018}
Minganti F, Biella A, Bartolo N and Ciuti C 2018 {\em Phys. Rev. A\/} {\bf
  98}(4) 042118

\bibitem{Albert_2014}
Albert V~V and Jiang L 2014 {\em Phys. Rev. A\/} {\bf 89}(2) 022118

\bibitem{Morley_2019}
Morley W~T, Di~Marco A, Mantovani M, Stadler P, Belzig W, Rastelli G and Armour
  A~D 2019 {\em Phys. Rev. B\/} {\bf 100}(5) 054515

\bibitem{Strogatz_book}
Strogatz S~H 1994 {\em Nonlinear Dynamics And Chaos: With Applications To
  Physics, Biology, Chemistry, And Engineering (Studies in Nonlinearity)\/}
  (Perseus Books)

\bibitem{Armour_2017}
Armour A~D, Kubala B and Ankerhold J 2017 {\em Phys. Rev. B\/} {\bf 96}(21)
  214509

\bibitem{QUTIP}
Johansson J, Nation P and Nori F 2013 {\em Computer Physics Communications\/}
  {\bf 184} 1234 -- 1240

\bibitem{Kenfack_2004}
Kenfack A and Zyczkowski K 2004 {\em Journal of Optics B: Quantum and
  Semiclassical Optics\/} {\bf 6} 396--404

\bibitem{Bartolo_2016}
Bartolo N, Minganti F, Casteels W and Ciuti C 2016 {\em Phys. Rev. A\/} {\bf
  94}(3) 033841

\bibitem{Roberts_2020}
Roberts D and Clerk A~A 2020 {\em Phys. Rev. X\/} {\bf 10}(2) 021022

\bibitem{Souquet_2016}
Souquet J~R and Clerk A~A 2016 {\em Phys. Rev. A\/} {\bf 93}(6) 060301

\bibitem{Schirmer_2010}
Schirmer S~G and Wang X 2010 {\em Phys. Rev. A\/} {\bf 81}(6) 062306

\bibitem{Macieszczak_2016}
Macieszczak K, Guţă M, Lesanovsky I and Garrahan J~P 2016 {\em Phys. Rev.
  Lett.\/} {\bf 116}(24) 240404

\bibitem{Kessler_2012}
Kessler E~M, Giedke G, Imamoglu A, Yelin S~F, Lukin M~D and Cirac J~I 2012 {\em
  Phys. Rev. A\/} {\bf 86}(1) 012116

\bibitem{Minganti_2020}
Minganti F, Arkhipov I~I, Miranowicz A and Nori F 2020 {\em arXiv:2008.08075\/}

\bibitem{QM_in_PS_book}
Curtright T~L, Fairlie D~B and Zachos C~K 2014 {\em A Concise Treatise on
  Quantum Mechanics in Phase Space\/} (World Scientific)

\bibitem{Groenewold_1946}
Groenewold H 1946 {\em Physica\/} {\bf 12} 405 -- 460

\bibitem{stroscio_dutta_2001}
Stroscio M~A and Dutta M 2001 {\em Phonons in Nanostructures\/} (Cambridge
  University Press)

\bibitem{Katarzyna_2020}
Macieszczak K, Rose D~C, Lesanovsky I and Garrahan J~P 2020 {\em
  arXiv:2006.01227\/}

\bibitem{Bauke_2011}
Bauke H and Itzhak N~R 2011 {\em arXiv:1101.2683\/}

\bibitem{Steuernagel_2013}
Steuernagel O, Kakofengitis D and Ritter G 2013 {\em Phys. Rev. Lett.\/} {\bf
  110}(3) 030401

\bibitem{Braasch_2019}
Braasch W~F, Friedman O~D, Rimberg A~J and Blencowe M~P 2019 {\em Phys. Rev.
  A\/} {\bf 100}(1) 012124

\bibitem{Dambach_2015}
Dambach S, Kubala B, Gramich V and Ankerhold J 2015 {\em Phys. Rev. B\/} {\bf
  92}(5) 054508

\bibitem{Svensson_2018}
Svensson I~M, Bengtsson A, Bylander J, Shumeiko V and Delsing P 2018 {\em
  Applied Physics Letters\/} {\bf 113} 022602 (\textit{Preprint}
  \eprint{https://doi.org/10.1063/1.5026974})

\bibitem{Brock_2020}
Brock B~L, Li J, Kanhirathingal S, Thyagarajan B, Jr W~F~B, Blencowe M~P and
  Rimberg A~J 2020 {\em arXiv:2011.06298\/}

\bibitem{Carmichael_book}
Carmichael H~J 1999 {\em Statistical Methods in Quantum Optics 1: Master
  Equations and Fokker-Planck Equations\/} (Springer)

\bibitem{mandel_wolf_book}
Mandel L and Wolf E 1995 {\em Optical Coherence and Quantum Optics\/}
  (Cambridge University Press)

\bibitem{FIPY}
Guyer J~E, Wheeler D and Warren J~A 2009 {\em Computing in Science \&
  Engineering\/} {\bf 11} 6--15

\bibitem{LANDAU_1999}
Landau L 1999 {\em Journal of Mathematical Analysis and Applications\/} {\bf
  240} 174 -- 204

\end{thebibliography}

\end{document}